\documentclass[journal]{IEEEtran}
\ifCLASSOPTIONcompsoc
  \usepackage[nocompress]{cite}
\else
  \usepackage{cite}
\fi

\usepackage{graphicx}
\usepackage{balance}
\usepackage{comment}
\usepackage{lscape}
\usepackage[normalem]{ulem}
\usepackage{booktabs}
\newcommand{\ra}[1]{\renewcommand{\arraystretch}{#1}}
\usepackage[ruled]{algorithm2e}
\usepackage{authblk}
\usepackage{verbatim} 
\usepackage{graphics} 
\usepackage{epsfig} 
\usepackage{epstopdf}
\usepackage{subfigure}
\usepackage{balance}
\usepackage{comment}
\usepackage{booktabs}
\usepackage{amsmath}
\usepackage{amssymb}
\usepackage{amsfonts}
\usepackage{bm}
\usepackage{colortbl}
\usepackage{tabularx}
\usepackage{color}
\usepackage{xspace}
\usepackage{graphicx}    
\usepackage{url}         
\usepackage{algpseudocode}
\usepackage{placeins}
\usepackage{multirow}
\usepackage{caption}
\usepackage{enumitem}
\usepackage{leading}
\usepackage{pgfplots}
\usepackage{tikz}
\usepackage{etex}
\usepackage{authblk}
\usepackage{caption}
\usepackage{collcell}
\usepackage{hhline}
\usepackage{pgf}

\newcommand\gray{gray}

\newcommand\ColCell[1]{%
	\pgfmathparse{#1<.8?1:0}%
	\ifnum\pgfmathresult=0\relax\color{white}\fi
	\pgfmathparse{1-#1}%
	\expandafter\cellcolor\expandafter[%
	\expandafter\gray\expandafter]\expandafter{\pgfmathresult}#1}

\newcolumntype{E}{>{\collectcell\ColCell}c<{\endcollectcell}}

\newcommand{\RN}[1]{%
	\textup{\uppercase\expandafter{\romannumeral#1}}%
}

\begin{document}

\title{ Sensing, Computing, and Communications for Energy Harvesting IoTs: A Survey\\}

\author{Dong Ma, \textit{Student Member,~IEEE,} Guohao Lan, \textit{Member,~IEEE,} Mahbub Hassan, \textit{Senior Member,~IEEE,} \hspace{1in} Wen Hu, \textit{Senior Member,~IEEE,} and Sajal K. Das, \textit{Fellow,~IEEE}
\thanks{Dong Ma is with the School of Computer Science and Engineering, University of New South Wales (UNSW), Sydney, Australia, and also with the Data61, CSIRO, Eveleigh, Australia (E-mail: dong.ma1@unsw.edu.au).}
\thanks{Guohao Lan is with the Department of Electrical and Computer Engineering, Duke University, Durham, NC 27708, USA (E-mail: guohao.lan@duke.edu). Most of the work was done while the author was with the University of New South Wales, Sydney, Australia.}
\thanks{Mahbub Hassan is with the School of Computer Science and Engineering, University of New South Wales (UNSW), Sydney, Australia (E-mail: mahbub.hassan@unsw.edu.au).}
\thanks{Wen Hu is with the School of Computer Science and Engineering, University of New South Wales (UNSW), Sydney, Australia (E-mail: wen.hu@unsw.edu.au).}
\thanks{Sajal K. Das is with the Department of Computer Science, Missouri University
of Science and Technology, Rolla, MO 65409 USA. He is also an International Visiting Professor and SERB-sponsored VAJRA Faculty at Indian Institute of Technology, Kharagpur, India (e-mail: sdas@mst.edu).}
}

\maketitle

\begin{abstract}
With the growing number of deployments of Internet of Things (IoT) infrastructure for a wide variety of applications, the battery maintenance has become a major limitation for the sustainability of such infrastructure. To overcome this problem, energy harvesting offers a viable alternative to autonomously power IoT devices, resulting in a number of battery-less energy harvesting IoTs (or EH-IoTs) appearing in the market in recent years. Standards activities are also underway, which involve wireless protocol design suitable for EH-IoTs as well as testing procedures for various energy harvesting methods. Despite the early commercial and standards activities, IoT sensing, computing and communications under unpredictable power supply still face significant research challenges. This paper systematically surveys recent advances in EH-IoTs from several perspectives. First, it reviews the recent commercial developments for EH-IoT in terms of both products and services, followed by initial standards activities in this space. Then it surveys methods that enable the use of energy harvesting hardware as a proxy for conventional sensors to detect contexts in energy efficient manner. Next it reviews the advancements in efficient checkpointing and timekeeping for intermittently powered IoT devices. We also survey recent research in novel wireless communication techniques for EH-IoTs, such as the applications of reinforcement learning to optimize power allocations on-the-fly under unpredictable energy productions, and packet-less IoT communications and backscatter communication techniques for energy impoverished environments. The paper is concluded with a discussion of future research directions. 
\end{abstract}


\section{Introduction}
With the advancements in low-power and miniature electronics, recent years have witnessed a dramatic increase of Internet of Things (IoTs) applications covering a wide range of application areas, such as civil infrastructure, home automation, consumer electronics, wearable devices, and industrial or agricultural monitoring, to name a few. This trend is unstoppable as we continue to automate processes in every sector of our economy and daily life. Forecasts suggest that by 2020, a massive 5.8 billion of IoT devices will be deployed worldwide~\cite{rivera2019gartner}.

As most of the IoT devices are designed as small, wireless portable consumer electronics, they are expected to be powered by \textit{on-device power supply}, which is currently realized by various types and sizes of batteries depending on the application requirements. Unfortunately, for large scale IoT deployments, the battery technology suffers from major limitations. Because batteries store a finite amount of energy, they either need to be recharged or replaced, which is not only inconvenient and costly, but also not possible in certain deployments. To prolong battery life, the IoT devices could be configured or scheduled to be active less frequently, at the cost of reduced utility. In some deployments, batteries may also pose huge safety risks. Finally, dumping billions of toxic batteries is not environment-friendly. Due to these reasons, powering a massive number of sensors is now recognized as one of the grand challenges of the IoT revolution~\cite{hester2017future,psikick}.

Harvesting energy from the ambient environment to perpetually power electronic sensors is a promising solution to eliminate dependency on batteries and thus accelerate the deployments of IoTs. Indeed, recent advancements in energy harvesting materials, devices, and processes have made it possible to realize certain IoT circuits that can operate without batteries. Examples of such energy harvesting IoTs (or EH-IoTs) include wireless switches \cite{WirelessSwitch}, which harvest kinetic energy from each push of the switch button to transmit a low-power wireless message to an actuator to turn on/off a globe or pull down/up a curtain, and so on. Other examples include smart shoes \cite{solepower,instep,powerWalk} harvesting energy from foot strikes, smartwatches powered by kinetic~\cite{sequent}, solar~\cite{lunar}, or thermal \cite{matrix} energy harvesting, energy meters \cite{debruin2013monjolo} powered by electromagnetic energy harvesting, and so on.  To promote interoperability between products from different manufacturers, the International Organization for Standardization (ISO) and the International Electrotechnical Commission (IEC) have initiated standardization of EH-IoTs, which involve suitable wireless protocol design as well as testing procedures for various types of energy harvesting methods. Such standardization activities are expected to further accelerate the development and deployment of EH-IoTs in the coming years.

\begin{figure}[t]
	\centering
	\centering
		\includegraphics[scale=0.45]{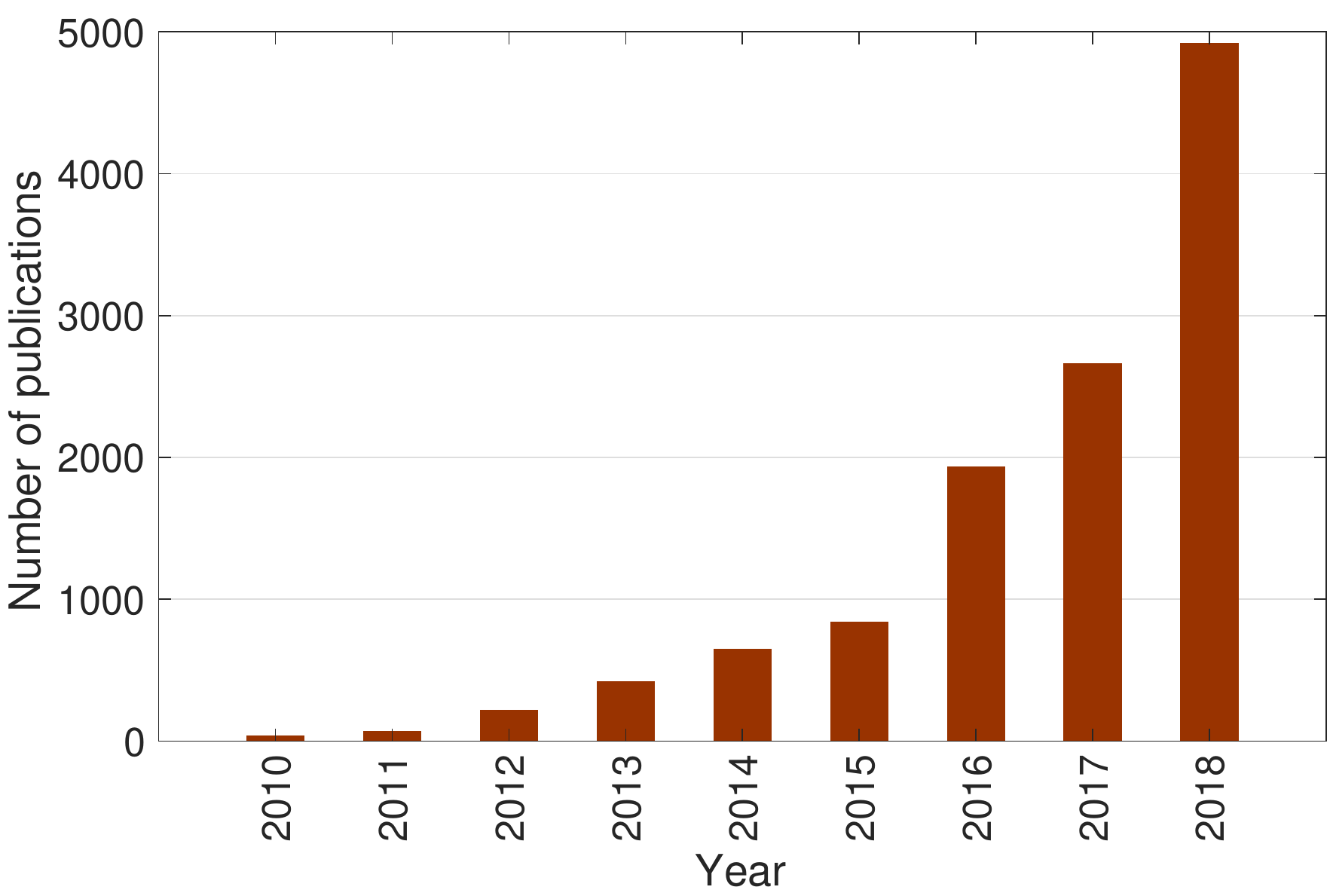}
	\caption{The exponential growth of the number of publications per year containing the following keywords: \textit{energy harvesting IoT, self-powered IoT, batteryless IoT, and battery-free IoT} (numbers are retrieved from \textit{Dimensions}\protect\footnotemark).}
	\label{fig:researchTrend}
\end{figure}

Despite the early commercial and standards activities, there remain significant research challenges to realize efficient and reliable EH-IoTs. For small form factor devices, current energy harvesting technology can produce only small amounts of power, which is also very dynamic and unpredictable. Continuously powering various sensors for 24/7 context monitoring using tiny amounts of harvested power is fundamentally challenging. Unpredictability of power generation causes further challenges for reliably completing various computing tasks and optimizing power allocation for wireless communications. How to optimize sensing, computing and communications for EH-IoTs has therefore become a topic of intense research. Figure \ref{fig:researchTrend} shows that the number of publications dealing with challenges in EH-IoTs have grown exponentially in recent years, confirming the popularity of the topic.  

\begin{table*}[t]
\centering
\caption{Comparison of existing EH surveys/reviews with this survey.}
\label{tab:survey_comp}
\setlength\tabcolsep{2.0pt}
\ra{1.3}
\begin{tabular}{ccccccccc}
\toprule

 & \multirow{2}{*}{\textbf{Energy Source}} & \multirow{2}{*}{\textbf{RF EH}} & \multirow{2}{*}{\textbf{Commercial} }& \multirow{2}{*}{\textbf{Sensing from}} & \multirow{2}{*}{\textbf{Intermittent} } & \multicolumn{3}{c}{\textbf{EH Communication}} \vspace{-0.04in} \\ \cmidrule{7-9}

\textbf{Paper} & \textbf{/Techniques} &\textbf{WPT\&SWIPT)} & \textbf{Standards/Activities }& \textbf{EH Patterns} & \textbf{Computing} & \textbf{EH-Optimized}&\textbf{Packet-less} &\textbf{Reflective}  \vspace{-0.03in}  \\

 &  &  &  & &  & \textbf{Transmission}&\textbf{Communication} &\textbf{Communication}\\ \hline
 
\cite{bhatti2016energy} & Comprehensive&Comprehensive &None & None  & None & None& None& None \\ \hline 
 
\cite{sudevalayam2011energy} & Comprehensive& None & None &None  & Brief Summary & Brief Summary& None& None \\ \hline 

\cite{lu2015wireless} & Moderate&Comprehensive& None & None  &None & Moderate& None& None \\ \hline 

\cite{he2015survey} & Brief Summary& None & None & None  & None & Moderate& None& None \\ \hline

\cite{ku2016advances} & Comprehensive& Moderate & None &None  & None & Comprehensive& None& None \\ \hline 

\cite{ahmed2015survey} & Brief Summary& None & None & None & None & Comprehensive& None& None\\ \hline 

\cite{ulukus2015energy}& Brief Summary&Moderate & None &None & None & Comprehensive& None& None \\ \hline 

\cite{kamalinejad2015wireless} & Moderate& Comprehensive & None &None  & None & None& None& None \\ \hline 

\cite{lucia2017intermittent} &Brief Summary& None & None &None  & Moderate & None& None& None \\ \hline 

\cite{khan2009survey}& None& Brief Summary & None & None  & None & None& None&RFID \\ \hline 

\cite{van2018ambient} & None & Moderate &Brief Summary &None  & None & Moderate& None& Ambient Backscatter \\ \hline 

\textbf{Ours} & \textbf{Brief Summary}& \textbf{None} & \textbf{Comprehensive} & \textbf{Comprehensive}  & \textbf{Comprehensive} & \textbf{RL-based} & \textbf{Comprehensive}& \textbf{IRS }\\ \hline 

\multicolumn{9}{l}{ *RL-based refers to Reinforcement Learning based online communication optimization, IRS refers to Intelligent Reflective Surface.}  
\end{tabular}
\end{table*}

\footnotetext{https://app.dimensions.ai/discover/publication}

\begin{figure}[t]
	\centering
	\centering
		\includegraphics[scale=0.5]{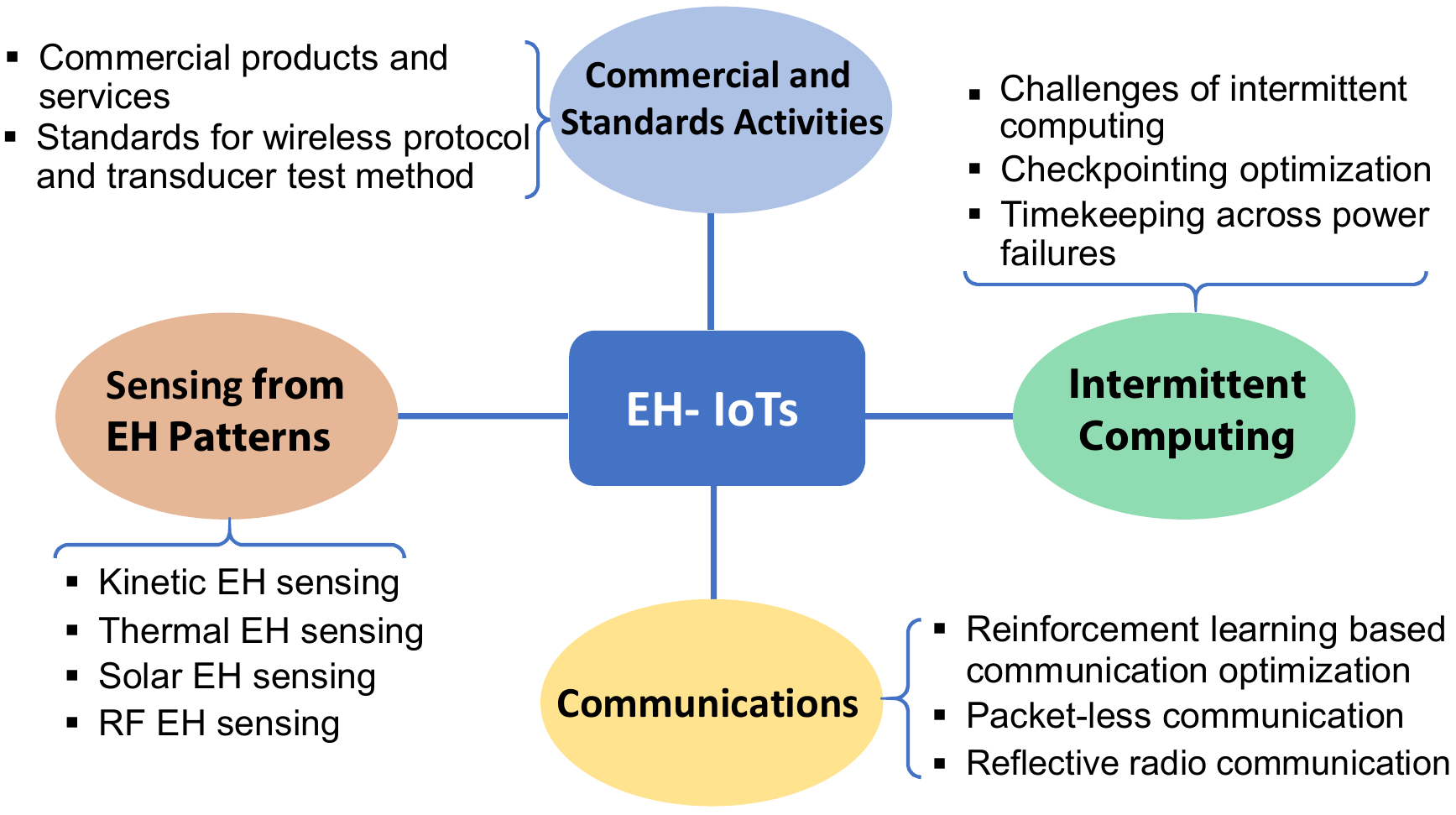}
	\caption{Taxonomy of our survey covering recent advances in commercialization, standardization, sensing, intermittent computing and communications for EH-IoTs.}
	\label{fig:overall taxonomy}
\end{figure}

The concept of energy harvesting is not new and there exist recent literature surveying recent advances in EH-related research. These surveys, however, do not capture many recent advances in EH-IoTs, such as those related to commercialization, standardization, context sensing from EH patterns, intermittent computing, or some emerging communication techniques. Our survey aims to fill this gap in the current literature. Table~\ref{tab:survey_comp} compares the coverage of our survey against previously published ones, while Figure~\ref{fig:overall taxonomy} shows a more detailed taxonomy of our survey. For topics extensively covered in the previous surveys, such as the energy harvesting techniques and transmission optimization solutions, we only provide a brief summary in this paper for the benefit of the readers and for the sake of completeness, while the details are referred to those references for additional information.

The survey is organized as follows. After a brief introduction to the basic architecture of an EH-IoT device, Section \ref{section:iots} surveys the recent commercial and standards developments in this area. Section \ref{section:sensing} surveys methods that enable use of energy harvesting hardware as a proxy for conventional sensors to detect contexts more efficiently in terms of energy.  The advancements on energy harvesting computing, including efficient checkpointing and timekeeping for intermittently powered scenarios are reviewed in Section \ref{section:computing}. Section \ref{section:communication} surveys recent research in novel wireless communication techniques for EH-IoTs, including applications of reinforcement learning to optimize power allocations on-the-fly under unpredictable energy productions as well as packet-less and backscatter communication of basic IoT notifications. Future research directions are envisaged in Section \ref{section:futurework}, before we conclude the survey in Section \ref{section:conclusion}.

\section{Commercial and Standards Developments in EH-IoTS}
\label{section:iots}
Although large scale pervasive deployments of EH-IoTs are yet to come, we are beginning to witness some early commercial successes in this area. There also exist fully working prototypes of EH-IoTs demonstrated in various research laboratories. Moreover, as summarized in Table~\ref{tab:EH companies1}, there is an upward trend in the number of third party vendors specializing in energy harvesting components and solutions suitable for IoTs. Collectively, these developments provide compelling evidence supporting the feasibility and promise of using renewable energy to self-power the next generation of IoTs. This section reviews recent developments in EH-IoTs including initial efforts for standardizing wireless communications suitable for EH-IoTs. To gain an insight to the working principle, we start this section by examining the system architecture of such devices.

\subsection{System Architecture of Emerging EH-IoTs}
\label{sec:self-power-architecture}

\begin{table*}[t]
\centering
\caption{Vendors specializing in energy harvesting components and solutions suitable for IoTs.}
\label{tab:EH companies1}
\setlength\tabcolsep{6.8pt}
\ra{1.3}
\begin{tabular}{lccclc}
\toprule

\textbf{Company} & \textbf{Source} & \textbf{Energy Harvester} & \textbf{EH Solution} & \textbf{Application} & \textbf{Foundation Year} \\ \hline
 
Piezo Systems & Kinetic& \checkmark & $\times$ &piezoelectric energy harvesting & 1988 \\ \hline 
MIDE Technology & Kinetic& \checkmark & $\times$ &piezoelectric energy harvesting & 1989 \\ \hline 
EnOcean & Kinetic/Solar & \checkmark & \checkmark  &Building & 2001 \\ \hline 
Perpetuum & Kinetic& $\times$ & \checkmark &Transportation  & 2004 \\ \hline 
Bionic Power & Kinetic& \checkmark & $\times$ & Wearables /Military & 2007 \\ \hline
PaveGen & Kinetic& \checkmark & \checkmark &Infrastructure /Entertainment & 2009 \\ \hline 
Eight19& Solar & \checkmark & \checkmark &Health care/Retail /Infrastructure  & 2010 \\ \hline 
SolePower& Kinetic & $\times$ & \checkmark &Wearables  & 2012 \\ \hline 
PsiKick& Kinetic/Solar/Thermal & $\times$ & \checkmark &Industry  & 2012 \\ \hline 
Greengineering& Thermal & $\times$ & \checkmark &Building/Industry/Transportation  & 2012 \\ \hline 
ReVibe Energy& Kinetic & \checkmark& \checkmark &Industry/Transportation  & 2013 \\ \hline 
Enerbee& Kinetic & $\times$ & \checkmark &Building/Industry  & 2014 \\ \hline 
AMPY& Kinetic & \checkmark & $\times$ &Wearables  & 2014 \\ \hline 
8power& Kinetic & $\times$ & \checkmark &Industry  & 2015 \\ \hline 
Freevolt& RF & \checkmark & $\times$ &Building/Wearables  & 2015 \\ \hline 
Nowi Energy& RF & \checkmark & \checkmark &Industry/Wearables/Transportation  & 2015 \\ \hline 
ParkHere& Kinetic & \checkmark & \checkmark &Infrastructure  & 2015 \\ \hline 
Aqua Robur& Kinetic & \checkmark & \checkmark &Industry/Irrigation  & 2015 \\ \hline 

Kinergizer& Kinetic & \checkmark & \checkmark &Industry/Wearables/Transportation  & 2016 \\ \hline 
Trameto& Multiple & $\times$ & $\times$ &Power management integrated circuits  & 2016 \\ \hline 
Otego& Thermal & \checkmark & \checkmark &Industry  & 2016 \\ \hline 
Zolitron Technology & Solar & \checkmark & \checkmark &Industry/Transportation  & 2016 \\ \hline 
Lightricity & Solar & \checkmark & $\times$ &Industry/Wearables/Building  & 2017 \\ \hline 
EnergIoT & Kinetic & \checkmark & \checkmark &Industry/Transportation  & 2017 \\ \hline
  
  
\end{tabular}
\end{table*}

  \begin{figure}[t]
  	\centering
  	\centering
 		\includegraphics[scale=0.5]{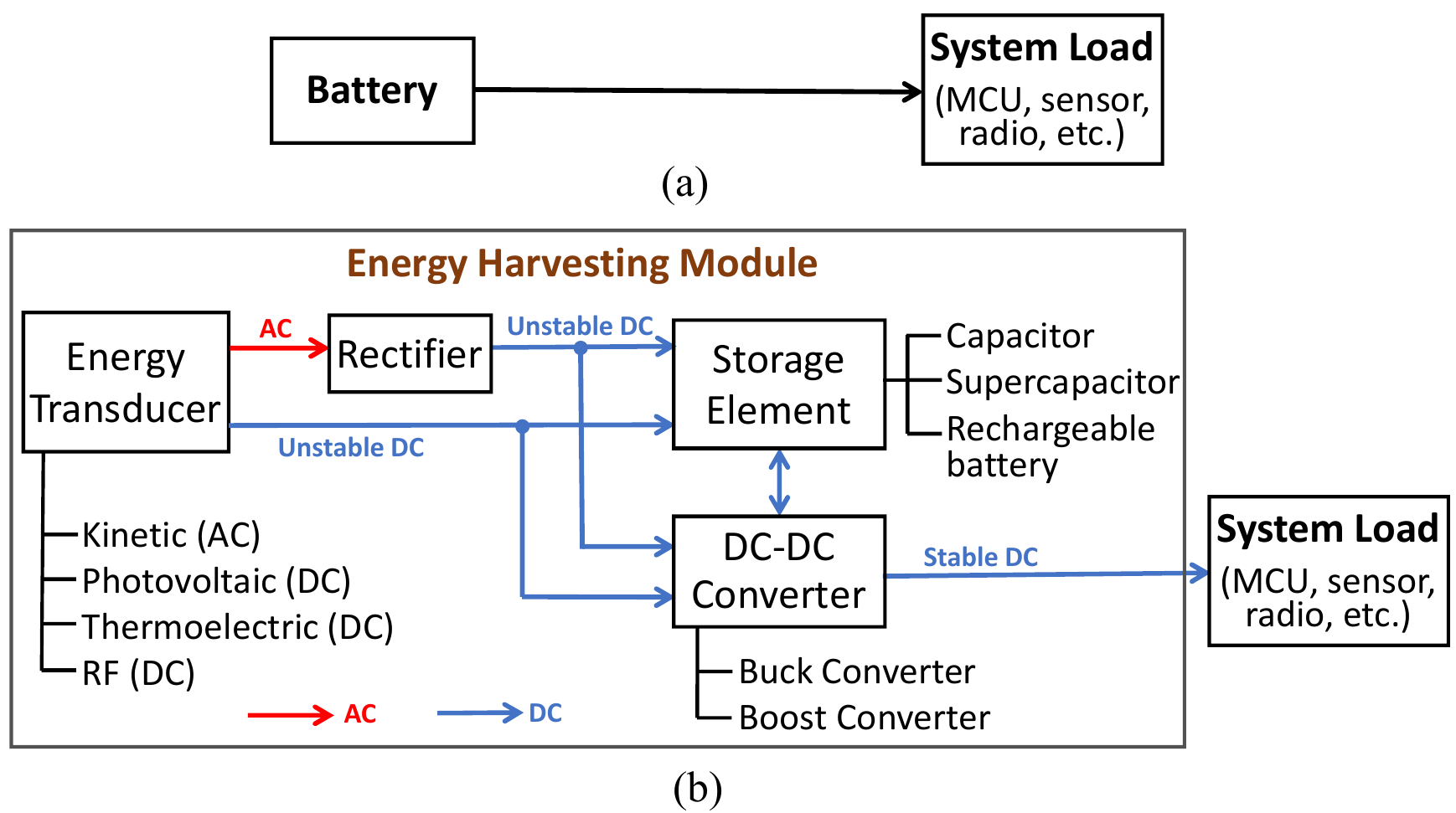}
  	\caption{Generic system architecture for (a) conventional IoT powered by battery and (b) emerging EH-IoT, which simply replaces the battery with an energy harvesting module. } 
	\label{fig:architecture}
  \end{figure}

It is important to design the architecture of EH-IoT to ensure existing electronic components (e.g., microprocessors, sensors, and radios) can be seamlessly reused in these new breed of devices. Although the actual details of the device architecture will vary from vendor to vendor, we capture the basic components and connectivity in Figure~\ref{fig:architecture}. As can be seen, the current design of EH-IoT allows us to simply replace the battery by an energy harvesting module, and the remaining IoT electronic components can be seamlessly integrated and powered by the module. This design simplicity is achieved by breaking the energy harvesting component into two subcomponents: an energy harvesting \textit{transducer} followed by a \textit{power management module}, which includes storage element, DC-to-DC converter, and rectifier if the power source produces alternating current (AC).

The fundamental purpose of the transducer is to convert ambient energy available in the environment in various forms into usable electricity that can be used to power the IoT electronics. Different transduction techniques and materials can be used to harvest energy from different sources. There are four main sources of ambient energy suitable for IoTs. They are kinetic (a.k.a motion, vibration, and mechanical), solar, radio frequency (RF), and thermal. Different sources produce energy under different contexts. For example, kinetic energy harvesting is only possible if the object moves, vibrates, bends, or experiences some type of motion. In contrast, a solar cell can produce electricity even for static objects, but it cannot work in the dark. A thermal source can generate electricity even for static objects in the dark as long as the temperature differences in space or time are present. Finally, the RF energy harvesting can convert ambient radio waves into usable electricity and would work for all type of objects as long as it is within the coverage of some type of radio transmitters. More details about the mechanisms, characteristics, and applications of these energy sources and the corresponding transduction techniques can be found in~\cite{chalasani2008survey,priya2009energy}; we compare and contrast them in Table \ref{t:sources} from the perspective of their use in EH-IoTs. As can be observed, given different application scenarios and requirements of EH-IoTs, we need to select the proper energy harvesting transduction techniques. For example, there is no point to employ solar energy harvesting if the object is expected to operate mostly in dark environments. 

\begin{table*}[t]
 \centering
 \caption{Typical energy harvesting sources and techniques used in EH-IoTs.}
 \label{t:sources}
 \ra{1.2}
 \setlength\tabcolsep{4.7pt}
 \begin{tabular}{llclll}
 \hline
 
 \textbf{Source} & \textbf{Transduction} & \textbf{Signal} & \multicolumn{1}{c}{\textbf{Principle}} & \textbf{Requirement} & \textbf{Energy Density} \\ \hline
  
  \multirow{6}{*}{Kinetic}& Piezoelectric & AC & Apply compression on crystalline materials &\multirow{6}{*}{Motions or vibrations} &Walking: $49 \mu W/cm^2$@$3km/h$,   \\ \cmidrule{2-4}
 
  & Electromagnetic & AC  & Change of magnetic field under movement & &  piezoelectric EH, knee bending~\cite{kuang2017energy} \\ \cmidrule{2-4}
 
  & Electrostatic & AC & Change of electrical field under movement & &Wind: $370 \mu W/cm^2$@$15m/s$, \\  \cmidrule{2-4} 
 
 &  Triboelectric & AC & Frictional contact of two different materials & & triboelectric EH~\cite{ahmed2017farms}\\ \cmidrule{1-6}
 
 \multirow{3}{*}{Thermal} & Thermoelectric & DC & Spatial temperature gradient &\multirow{3}{*}{Temperature difference} & $14uW/cm^2$ on human wrist,\\ \cmidrule{2-4}
 
  & Pyroelectric &AC & Temporal temperature difference & & thermoelectric EH~\cite{thielen2017human} \\ \cmidrule{1-6}
 
 \multirow{3}{*}{Solar} & \multirow{3}{*}{Photovoltaic} &\multirow{3}{*}{DC}  & \multirow{3}{*}{Convert light into electricity} &\multirow{3}{*}{Bright environment} & Transparent:$7mW/cm^2$@128klux~\cite{upama2017high}    \\ 
 
 &&&&& Opaque: $26.7mW/cm^2$@128klux~\cite{yoshikawa2017silicon}\\
 
 &&&&& Opaque: $16 \mu W/cm^2$@400lux~\cite{lucarelli2017efficient}\\ \cmidrule{1-6}

 \multirow{2}{*}{RF} & \multirow{2}{*}{RF radiation} & \multirow{2}{*}{AC} & \multirow{2}{*}{Convert electromagnetic wave into electricity} & \multirow{2}{*}{Radio coverage} &TV tower:$60 \mu W$@distance 4.1km~\cite{sample2009experimental}    \\ 
 
 &&&&& Wi-Fi: $100\mu W$@distance 2 feet~\cite{talla2015powering}\\ 
 
 \hline   
 \end{tabular}
 \end{table*}

The transducer, however, can only generate electricity as energies become available in the environment, which means that the transducer output is highly unstable and cannot be used directly to power conventional electronic components that are designed for stable direct current (DC) supply. A power management module is therefore required to further regulate the electricity generated by the transducer in order to produce stable power supply for the IoT electronics. Such regulation may involve simple DC-to-DC conversion, or if the transducer produces AC, it may also require AC-to-DC conversion (i.e., rectifier) as part of the energy harvesting regulation. Finally, as electricity generation can be very dynamic and intermittent, it is often necessary to store the harvested electricity for a while before using it. Therefore, the power management module also often includes a capacitor or a rechargeable battery.

\subsection{Commercial and Academic EH-IoTs}
With the efforts of researchers and engineers, there exits a wide range of self-powered IoT devices in the academia and commercial sectors. These devices achieve energy autonomous operation merely relying on the energy harvested from the environment. The application domain varies from smart building and transportation, to wearable and implantable medical devices. In this subsection, we review the existing battery-free prototypes or products in different application domains. 

\begin{table*}[t]
\caption{Examples of self-powered IoTs: commercial products and research prototypes.}
\label{tab:self-powered IoTs}
\ra{1.2}
\setlength\tabcolsep{2.2pt}
\begin{tabular}{llllccllll}
\hline
\multicolumn{1}{c}{\multirow{2}{*}{\textbf{Domain}}} & \multicolumn{1}{c}{\multirow{2}{*}{\textbf{Product}}} & \textbf{EH} & \multirow{2}{*}{\textbf{A/C}}  & \multirow{2}{*}{\textbf{Radio}} & \textbf{Release}  & \textbf{Energy}& \multirow{2}{*}{\textbf{Function}} & \textbf{Energy Harvesting Capability} \\ 

\multicolumn{1}{l}{} & \multicolumn{1}{c}{}  & \textbf{Technique} & \textbf{} & & \textbf{Year} & \textbf{Storage} &&\textbf{(Form Factor)}\\ \hline

& \multirow{2}{*}{Sensors~\cite{tem_hum}} &\multirow{2}{*}{Solar/Photovoltaic} & \multirow{2}{*}{C} 
& \multirow{2}{*}{863MHz} & \multirow{2}{*}{-} & \multirow{2}{*}{Capacitor}   & Temperature &5min $@400$lux to send the first telegram  \\

& &  &  
&  &  &    & monitoring  &from cold start (solar cell: 50$\times$20mm) \\ \cline{2-9}

& Wireless  & \multirow{2}{*}{Kinetic/Piezoelectric} & \multirow{2}{*}{C} 
& \multirow{2}{*}{868MHz}& \multirow{2}{*}{-} & \multirow{2}{*}{-}  & Dimming/shutter   &Each push and release of button can actuate  \\

& Switch\cite{WirelessSwitch} &  &  
& &  &   &  control & a telegram transmission \\ \cline{2-9}

Smart& CleanSpace  & \multirow{2}{*}{Radio frequency} & \multirow{2}{*}{C}
& \multirow{2}{*}{BLE} & \multirow{2}{*}{2015} & \multirow{2}{*}{- }& Air quality         &   \multirow{2}{*}{-}  \\ 

&  Tag\cite{CleanSpaceTag} &   & 
&  &  &  &  monitoring &\\ \cline{2-9}

Building& \multirow{2}{*}{\cite{li2018trinity}}& \multirow{2}{*}{Kinetic/Piezoelectric} & \multirow{2}{*}{A}  
& \multirow{2}{*}{Zigbee}  & \multirow{2}{*}{2013}   & \multirow{2}{*}{Battery} & Air conditioning   & Generate 667$\mu$W at 5.5m/s airflow speed     \\

& &  &   
&   &    &  &  monitoring  & (piezoelectric EH: 20cm$^2$)      \\ \cline{2-9}

& Monjolo~\cite{debruin2013monjolo} & Electromagnetic & A 
& Zigbee   & 2013    & Capacitor & Energy metering      &   Generated 4mW at 60W load     \\ \cline{2-9}
& WISP~\cite{sample2008design} & Radio frequency & A 
& 928MHz   & 2008    & Capacitor & Context monitor    &Generate 310$\mu$W with 0dBm input power          \\ \cline{2-9}
& EnHANT~\cite{gorlatova2013networking} & Solar/Photovoltaic & A 
& UWB-IR   & 2013    & Battery & Object tracking  &Generated 70$\mu$W/cm2 at lab environment            \\ \cline{1-9}
& Lunar    & \multirow{2}{*}{Solar/Photovoltaic}  & \multirow{2}{*}{C}  
& \multirow{2}{*}{BLE}  &\multirow{2}{*}{2017}    & \multirow{2}{*}{Battery} & Step counting,  & Exposure at $>10$K lux for 1 hour can  \\

& Watch\cite{lunar}    &   & 
&   &    &  & Sleep tracking & support 24 hours operation \\ \cline{2-9}

& Sequent   & Kinetic/  & \multirow{2}{*}{C} 
& \multirow{2}{*}{BLE}  & \multirow{2}{*}{2017}    & \multirow{2}{*}{Battery} & Activity, heart   &  -    \\

& Watch\cite{sequent}    & Electromagnetic  &  
&   &     &  &  rate tracking   &     \\ \cline{2-9}

& \multirow{2}{*}{PowerWatch\cite{matrix}}  & Thermal/  & \multirow{2}{*}{C}
& \multirow{2}{*}{BLE}  & \multirow{2}{*}{2017}    & \multirow{2}{*}{Battery} & Step counting,   & Can operate using the energy converted    \\

&  & Thermoelectric  & 
&   &     &  & sleep tracking  &    from body heat   \\ \cline{2-9}

& \multirow{2}{*}{Smartboots~\cite{solepower}}  & Kinetic/  &\multirow{2}{*}{C}  
& Wi-Fi/  & \multirow{2}{*}{2012}  & \multirow{2}{*}{-} & Wearable and   &  Generate 100mW    \\

&   & Electromagnetic  &  
& Cellular  &   &  & military motoring &  (form factor: shoe)     \\ \cline{2-9}

Wearable&INSTEP~\cite{instep}  & Kinetic  & C  
& BLE  & 2011  & Battery & Activity tracking,   & Generate 1W@walking (form factor: shoe)   \\ \cline{2-9}
&\multirow{2}{*}{PowerWalk~\cite{powerWalk}}  & \multirow{2}{*}{Kinetic/Piezoelectric}  & \multirow{2}{*}{C}  
& \multirow{2}{*}{-}  & \multirow{2}{*}{2007}  & \multirow{2}{*}{-} & Wearable and & Generate 11W from knee bending  \\ 

&  &   &   &   &   &  & military motoring & (form factor: knee brace)      \\ \cline{2-9}
& \multirow{2}{*}{AMPY~\cite{ampy}}  & Kinetic/  & \multirow{2}{*}{C}
& \multirow{2}{*}{BLE}  & \multirow{2}{*}{2014}  & \multirow{2}{*}{Battery} & \multirow{2}{*}{Power charger}  &Generate milliwatts from activity        \\ 

&  & Electromagnetic  & 
&   &   &  & &(form factor: power bank)      \\\cline{2-9}

&\multirow{2}{*}{Sunglasses\cite{landerer2017solar}} & \multirow{2}{*}{Solar/Photovoltaic}  &\multirow{2}{*}{A}  
& \multirow{2}{*}{-}  & \multirow{2}{*}{2017}  & \multirow{2}{*}{Capacitor} & Temperature and  & Generate 400$\mu$W at 500lux \\ 

& &   &   
&  &   &  & light monitoring & (form factor: glass lens, solar cell 31cm$^2$) \\ \cline{2-9}

&\cite{huang2018toward} & Kinetic/Piezoelectric  & A  
& BLE  & 2016  & Capacitor & Activity tracking & Generate 1-2mW (piezoelectric EH: 50cm$^2$)  \\  \cline{1-9}

&\multirow{2}{*}{HiPER-D\cite{kinergizer}} & Kinetic/ & \multirow{2}{*}{C} 
& \multirow{2}{*}{-} & \multirow{2}{*}{2016} & \multicolumn{2}{l}{Energy harvesters tha} &Generate 3mW at vibration of 1g and 20Hz \\ 

& & Electromagnetic &  
&  &  & \multicolumn{2}{l}{ support customized} &(cylinder: height 75mm and diameter 35mm) \\ \cline{2-6} \cline{9-9}

Industry&Model  & Kinetic/ & \multirow{2}{*}{C} 
& \multirow{2}{*}{-} & \multirow{2}{*}{2013} &\multicolumn{2}{l}{ asset condition} & Generate 150mW at vibration of 1g and 60Hz \\ 

& A/D/Q\cite{revibe} & Electromagnetic &  
&  &  &\multicolumn{2}{l}{ monitoring, tracking, etc.} & (form factor: power bank, cylinder battery)\\\cline{2-9}

&Fenix Hub\cite{fenixhub} & Kinetic/turbine & C 
& LoRa & 2015 & Battery & Irrigation & Generate up to 4W (200$\times$160$\times$65mm) \\  \cline{2-9}
&\multirow{2}{*}{Z-Node\cite{znode}} & \multirow{2}{*}{Solar/Photovoltaic} & \multirow{2}{*}{C} & BLE/ & \multirow{2}{*}{2016} & \multirow{2}{*}{Battery} &Asset tracking & \multirow{2}{*}{(form factor: 80$\times$90$\times$12mm)}   \\ 

& & & & NB-IoT& & &Smart traffic& \\ \cline{1-9}
& Pacemaker\cite{deterre2014micro} & Kinetic/Piezoelectric  & A
& -  & 2014  & - & Drive pacemaker  & Generate 3 $\mu$J/cm$^3$/heartbeat(Cylinder 21mm$^3$)      \\  \cline{2-9}
IMD& Heart monitor\cite{zheng2016vivo}  & Kinetic/Triboelectric  & A
& -  &2016   & Capacitor & Heartbeat monitor  &   Generate 19.5nW at heart beat of 80 bpm   \\ \cline{2-9}

& \multirow{2}{*}{Pacemaker\cite{southcott2013pacemaker}}  & \multirow{2}{*}{Biofuel/Glucose}  & \multirow{2}{*}{A}
& \multirow{2}{*}{- } & \multirow{2}{*}{2013}  & \multirow{2}{*}{-} &   \multirow{2}{*}{Drive pacemaker} &Generate 470mV voltage, 5mA current       \\ 
&   &  & &   &   &  &  &(form factor: 10$\times$4$\times$0.3cm)    \\ 

\hline 

\multicolumn{8}{l}{ *A/C represents Academia prototype or Commercial product.}

\end{tabular}
\vspace{-0.2in}
\end{table*}

\textit{1) \textbf{Smart building}:} A smart building deploys many different IoT devices, such as temperature or humidity sensors, smoke detectors, camera sensors (for surveillance) and wireless switches, to enable a smart and convenient control of the environment. However, the deployment and maintenance of large number of IoT devices incur high cost including labor. Meanwhile, indoor environment provides a variety of energy sources like light, radio frequency (e.g., Wi-Fi), and kinetic energy from interaction with human beings. Therefore, collecting such energy to power IoT devices would be possible to reduce the cost due to deployment (by eliminating wires) and maintenance (by eliminating the replacement of battery). Many prototypes and products in this category exist in academia and commercial arena.

In~\cite{li2018trinity}, the authors presented a self-powered airflow sensing and control system. By mounting the piezoelectric energy harvester to the outlets of the air conditioning system, it can harvest energy from the airflow and power the sensor nodes which report the airflow speed to a server, thereby achieving real-time monitoring and control. In~\cite{debruin2013monjolo}, a batteryless power meter named Monjolo was developed. By attaching an electromagnetic energy harvester to the power line, based on the Faraday's law, electricity was generated when the current flows through. Once the harvested energy reaches a certain value, it was used to transmit a pulse, and eventually, the power load can be estimated based on the pulse reception frequency/rate. Using the electromagnetic energy harvester, the authors in~\cite{chiang2018tethys} also designed a batteryless sensor, called Tethys, which harvests energy from water flows. The harvested energy was leveraged to transmit the time stamp of the start and end of a shower process, thereby estimating the amount of water consumed. The authors in~\cite{kashimoto2016low} proposed an in-home living activity recognition system which utilizes passive infra-red sensors and door sensors to detect different activities, like eating, sleeping, cooking, working and so forth. The sensors are equipped with solar panels to harvest energy from sunlight or bulb light. 

In the commercial market, there also exist some self-powered IoTs for smart building. For example, the Wireless Switch~\cite{WirelessSwitch} from EnOcean is able to transmit a radio telegram for remote monitoring of switch status when it is pushed every time. It also provides convenient control of lighting, temperature and miscellaneous electric loads. Similarly, CleanSpace Tag~\cite{CleanSpaceTag} is a batteryless air pollution monitor that harvests energy from the ambient RF signals radiated by Wi-Fi access points or cellular base stations. It can track air pollution condition and record the users' exposure to harmful carbon monoxide so that it protects them from harmful air pollutions.

WISP~\cite{sample2008design} is a RF powered IoT device with capability of environmental sensing, and has been widely used in academic research. It has two descendant versions, Moo~\cite{zhang2011moo} and SocWISP~\cite{yeager2010}, which are upgraded in term of microcontroller, flash memory, and software firmware. Another example is EnHANT Tags~\cite{margolies2015energy} which harvest energy from indoor light using organic solar cells and exchange tag IDs with the neighboring Tags through UltraWideband Impulse Radio (UWB-IR) for object tracking, for example, locating a misplaced book in a library~\cite{gorlatova2010energy}.

\textit{2) \textbf{Wearable devices}:} Harvesting energy from human activities or body heat are viable solutions to power wearable devices. In~\cite{huang2018toward}, the authors proposed a shoe based battery-free wearable sensing platform where the power is generated from the two feet when people walk. The designed piezoelectric energy harvester achieved a power output of 1-2$mW$, which enables the operation of sensors, microcontroller unit (MCU), and radio with reasonable duty-cycling. Similarly, the authors in~\cite{chamanian2016wearable} presented a batteryless sensor platform which harvests energy from human motions. The sensor node equipped with an electromagnetic energy harvester was able to adjust its sensing rate and data transmission rate according to the jogger's activity. In~\cite{nicosia2018efficient} was developed SensorTile, a self-powered wristband, which is integrated with three photovoltaic strips. Combining with the on-board sensors, the wristband was able to detect human activities and communicate with a smartphone using Bluetooth Low Energy (BLE). Similarly, a sunglass that is able to measure the sunlight intensity and ambient temperature without the need of battery was developed in~\cite{landerer2017solar}. The two lenses of the sunglass were fitted with semi-transparent solar cells, which harvests energy from the sunlight. 

Some self-powered wearable products are also available in the market. A smart shoe from Instep NanoPower~\cite{instep} enables human activity recognition, where the energy is harvested from daily walking and running. There are several smartwatches which can operate batteryless by harvesting energy from human body. The Sequent Watch~\cite{sequent} is equipped with an electromagnetic energy harvester and relies on human wrist motions to power the functions. Matrix PowerWatch~\cite{matrix} harvests energy from body heat through a thermoelectric energy harvester, while the Lunar Watch~\cite{lunar} equipped with transparent solar cells harvests energy from the sunlight. All such smartwatches support popular wearable applications like step counting and health monitoring.

\textit{3) \textbf{Industry/transportation}: } Self-powered IoT devices have also been deployed in transport and industrial domains. Different motion energy harvesters are designed to cater for a variety of applications, such as process control, asset condition monitoring, tracking, and railway condition monitoring. Examples like HiPER from Kinergizer~\cite{kinergizer} as well as modelA/modelD from ReVibe Energy~\cite{revibe}, have been adopted in practical scenarios. For instance, when HiPER is mounted on the lower flange of the track, it is able to harvest energy from passing trains and power the integrated sensors for monitoring services, like condition monitoring of wheels and bearings. Similarly, in~\cite{vijayaraghavan2010novel}, the authors presented a batteryless wireless sensor node equipped with a piezoelectric energy harvester to collect energy from vibrations caused by passing vehicles. The harvested energy was used to power the radio for data transmission so that the remote server was able to analyze the traffic flow.

\textit{4) \textbf{Implantable medical devices}:} In the past decades, implantable medical devices (IMDs) have been designed and implemented to observe human physical actions, enhance the functionality of some damaged or degraded organs, and deliver drugs for the therapy of special diseases. Various medical devices, such as cardiac pacemakers, cochlear implants, tissue stimulator, and so on~\cite{dakurah2015implantable,bhunia2015implantable} have been widely used to provide physical treatment as well as assist healthcare tracking services in clinical practices.

Given that reliability is a critical concern, IMDs typically rely on batteries to maintain sustainable operations. However, one major drawback of battery-powered IMDs is its limited lifespan with only several years~\cite{bock2012batteries}, which often require battery replacement through surgery. Therefore, research exploring energy harvesting IMDs has been conducted in recent years.

In~\cite{lu2015ultra}, energy harvested from heart motions of a swine was collected using an implanted piezoelectric device and the results showed the feasibility to drive artificial pacemakers.
A triboelectric generator with output voltage up to 14$V$ and output current up to 5$\mu A$ was designed and fabricated in~\cite{zheng2016vivo}. Implanting the generator into an adult swine for over 72 hours, the results demonstrated that it can achieve self-powered wireless transmission for real-time heartbeat monitoring. In~\cite{southcott2013pacemaker}, the authors used a single biofuel cell to harvest energy from human body and designed a power management module to activate a pacemaker. The results show that it is able to power the pacemaker for at least 5 hours.  

To conclude, self-powered IoTs have already started appearing in several application domains and new companies continue to innovate, design, and market novel energy harvesting solutions for IoTs. From Table~\ref{tab:self-powered IoTs}, we can observe that kinetic energy gains most popularity for wearables, industry and IMDs, while smart building has the opportunity to adopt a wide range of energy sources.

\begin{table}[t]
 \centering
 \caption{Standardization activities for EH-IoTs.}
 \label{t:standardization}
 \setlength\tabcolsep{2.8pt}
 \ra{1.2}
 \begin{tabular}{cllll}
 \hline
 
 \textbf{Aspect} & \textbf{Type} & \textbf{Standard} & \textbf{Y}& \textbf{Domain} \\ \hline
  
 & Amplitude  & ISO/IEC  & \multirow{2}{*}{2012} &      \\ 
 
 Wireless& modulation & 14543-3-10~\cite{14543-3-10} &  & Home    \\ \cmidrule{2-4}

  Protocol & Frequency  & ISO/IEC  & \multirow{2}{*}{2016} & electronic  \\ 
 &  modulation &  14543-3-11~\cite{14543-3-11} &  &   \\ \cmidrule{1-5}
 
 \multirow{13}{*}{Transducer}& \multirow{4}{*}{Piezoelectric} & IEC 62969-3~\cite{62969-3} & 2018 &  Vehicle sensor   \\ \cmidrule{3-5}
 
 & & IEC 62830-4~\cite{62830-4} & 2018 &     \\ \cmidrule{3-4}
 
 & & IEC 62830-1~\cite{62830-1} & 2017 &    \\ \cmidrule{2-4}

EH & \multirow{3}{*}{Electromagnetic} & IEC 62830-3~\cite{62830-3} & 2017 &  \multirow{2}{*}{Consumer/}   \\ \cmidrule{3-4}
 
 &  & IEC 62407-28~\cite{62407-28} & 2017 & \multirow{2}{*}{Military/}   \\ \cmidrule{2-4}
 
 &\multirow{3}{*}{Thermoelectric} & IEC 62830-2~\cite{62830-2} & 2017 & \multirow{2}{*}{Industry}   \\ \cmidrule{3-4}
 
 & & IEC 62830-5~\cite{62830-5} & 2018 &    \\ \cmidrule{2-4}
 
 &\multirow{3}{*}{Triboelectric} & IEC 62830-6~\cite{62830-6} & 2018 &   \\ \cmidrule{3-4}
 
 & & IEC 62830-7~\cite{62830-7} & 2018 &     \\

 \hline
 \multicolumn{4}{l}{ *Y represents Release Year.}
   
 \end{tabular}
 \end{table}

\subsection{Standardization Activities for EH-IoTs}

To promote interoperability, the International Organization for Standardization (ISO) and the International Electrotechnical Commission (IEC) have taken initial steps to standardize two important aspects of EH-IoT, namely wireless communication protocols and transducer testing methods. In the following we describe these activities, a summary of which is provided in Table \ref{t:standardization}.

\textit{1) \textbf{Standardization of wireless protocols}:} This activity seeks to standardize new wireless protocols that are suitable for energy harvesting devices, which have access to extremely small amount of energy and may not have consistent supply of energy. The ISO and IEC have jointly released a standard for wireless short packet (WSP) protocol optimized for energy harvesting. This protocol is targeted for smart home applications such as lighting, heating, energy management, blinds control, different forms of security control and entertainment (audio and video) where various sensors and switches transmit very short command and control messages. The WSP is designed to carry such short messages using minimal number of bits to improve the chances of successful transmission even if the amount of harvested energy is extremely small. 

The OSI layers 1-3 have been specified and the medium access control (MAC) does not enforce the conventional carrier sensing or listen before talk (LBT) prior to transmission. Instead, devices are allowed to transmit on the channel straightaway whenever they want to, which is referred to as \textit{random access} within the standard. The random access mechanism serves two important functions. First, devices equipped only with transmitters but no receivers, can still effectively participate in the IoT eco-system. Second, even if the device has a receiver, it can choose to turn it off when transmitting if the harvested energy cannot power both the transmitter and receiver at the same time. To avoid collisions, the standard recommends such devices to operate at very low duty cycling or implement retransmissions at higher layers. 

Two different modulation schemes, amplitude modulation (AM) and frequency modulation (FM), have been proposed for WSP to deal with both energy efficiency and mobility of IoTs. The AM is more energy efficient, but less effective for mobile objects because the antenna impedance gets affected when held in hand or placed on metal surfaces. It also affects the amplitude linearity but not the frequency. Therefore, FM is recommended for mobile devices. While low frequencies, such as 315 MHz, can be used for AM, frequencies above 800 MHz is recommended for FM communications to achieve a reasonably small size for the antenna which suits mobile devices.

\textit{2) \textbf{Standardization of test methods for energy harvesting transducers}:} This activity seeks to devise standard methods for testing and evaluating specific types of transducers, such as those used for vibration and thermal energy harvesting. The IEC released nine standards so far that cover testing methods for piezoelectric, electromagnetic, thermoelectric, and triboelectric energy harvesting. The applications of these transducers involve consumer, military, and industrial electronics.

\begin{figure}
	\centering
	\centering
fig	\includegraphics[scale=0.65]{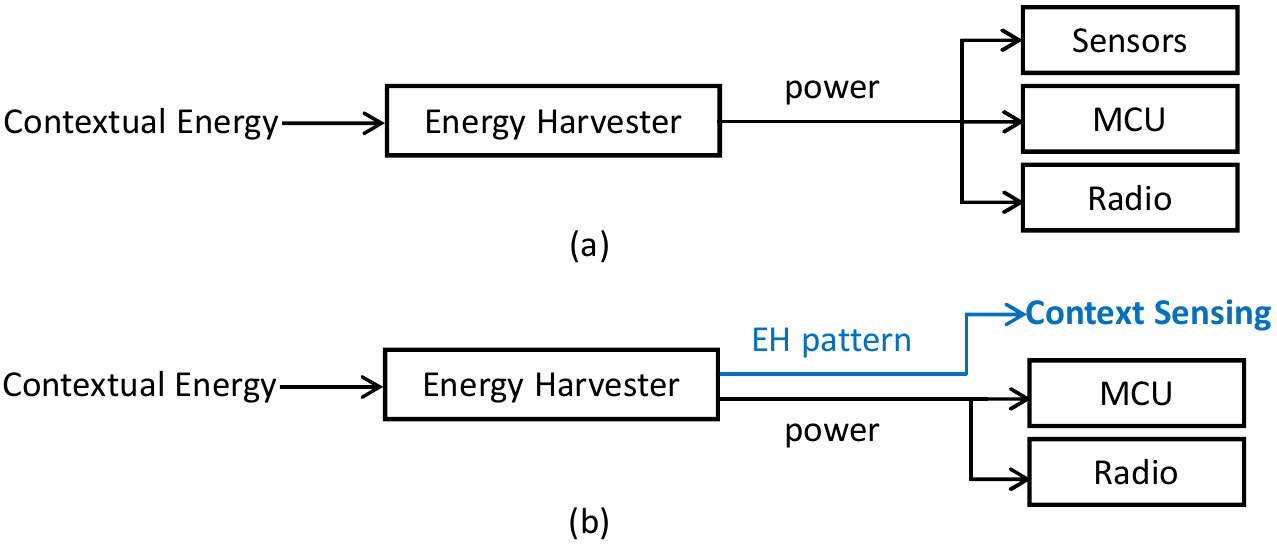}
	\caption{Paradigm of (a) current EH-IoTs and (b) novel EH-IoTs using energy harvesters as sensors.}
	\label{fig:sensorless_sensing}
\end{figure}

These initial standardization activities are expected to accelerate the development and deployment of energy harvesting IoTs in the coming years. Despite such commercial and standards activities, there remain plenty of opportunities to further optimize the sensing, computing, and communications tasks of an EH-IoT, which has become a topic of intense research in recent years. We survey these research works in the rest of this paper.

\begin{figure}
	\centering
	\centering
	\includegraphics[scale=0.57]{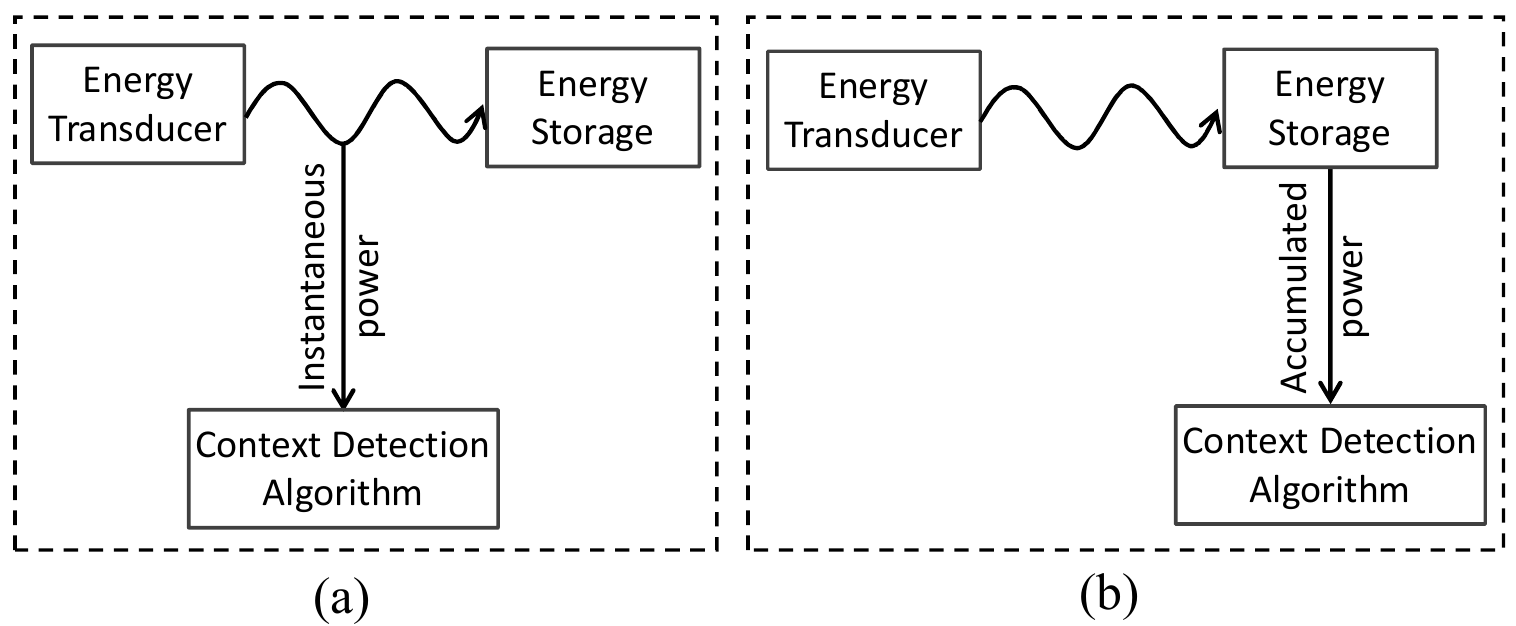}
	\caption{Illustration of using (a) Instantaneous and (b) Accumulated power signal for context sensing.}
	\label{fig:sensing taxonomy}
\end{figure}

\section{Context Sensing from EH Patterns}
\label{section:sensing}

Sensing various user and environmental contexts is the main application for many IoTs. Conventionally, context detection tasks are achieved with the help of specialized sensors, such as inertial sensor, microphone, GPS, light sensor, and so on. These specialized sensors, however, require external power supply to operate, making their extended use problematic for EH-IoTs, especially during energy-starving periods. Fortunately, based on the observation that energy harvesting patterns often reflect the context in which energy is being harvested, many researchers have successfully demonstrated the detection of a wide variety of contexts by reusing kinetic, thermoelectric, solar, and RF energy harvesting patterns. This section surveys these recent studies, a summary of which is presented in Table~\ref{tab:sensing-application-overall}.

\subsection{Energy Harvesters as Sensors}

As shown in Figure~\ref{fig:sensorless_sensing}(a), to sense various contexts, current EH-IoTs utilize the harvested energy to power sensor, MCU, and radio. To maintain continuous monitoring services, the energy consumption of sensors significantly affects the operation lifetime, especially in energy-constrained environment. As shown in Figure~\ref{fig:sensorless_sensing}(b), EH-IoTs also open up a new opportunity for context sensing by reusing the energy harvesting signals. For example, kinetic-powered wearable IoTs are able to detect and count the user's step, as the energy harvester generates distinguishable peaks in the energy harvesting signal each time the legs hit the ground \cite{khalifa2015step}. Similarly, a thermoelectric energy harvester is able to detect any changes in surface temperature simply from the variations in the generated energy harvesting signal~\cite{martin2012doubledip,campbell2014energy}. These examples encourage the use of energy harvesters as \textit{self-powered sensors} for EH-IoTs, which yields tangible power-saving opportunities as well as simpler and more compact hardware. 

In recent years, the EH patterns from kinetic, thermoelectric, solar, and RF energy harvesters have been demonstrated to detect a variety of contexts. Irrespective of the type of energy source, there are two main approaches for sensing from energy harvesting signals as illustrated in Figure~\ref{fig:sensing taxonomy}. The first approach analyzes the patterns of the \textit{instantaneous power} generated by the energy harvesting transducer, while the second approach uses the amount of the \textit{total energy accumulated} in the storage over a specific period of time. The first approach allows the detection of a rich set of contexts at the expense of more frequent sampling of the fluctuating power values. In contrast, the second approach can offer more significant power saving by sampling the stored energy only once in a while at the expense of more coarse-grained context sensing. Next, we review recent research works that apply these two approaches for context sensing with different EH schemes.

\subsection{Context Sensing from Kinetic EH}

\begin{figure}
	\centering
	\subfigure[]{
		\includegraphics[scale=0.82]{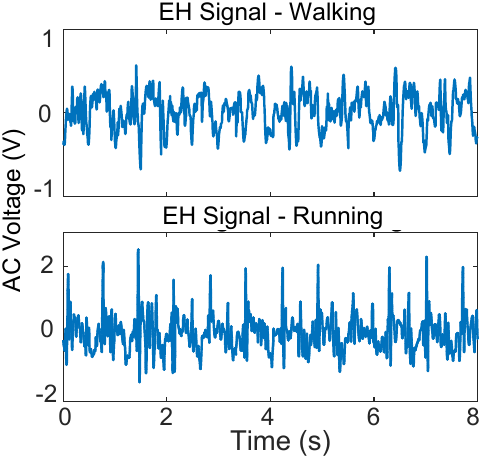}
		\label{fig:inst_signal_illustration}
	}
	\subfigure[]{
			\includegraphics[scale=0.82]{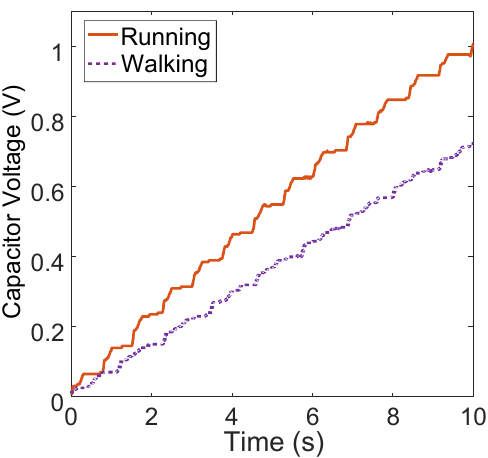}
			\label{fig:accu_signal_illustration}
		}
	\caption{Illustration of (a) instantaneous AC voltage and (b) accumulated capacitor voltage when a subject is walking and running wearing a piezoelectric energy harvester. The patterns of the AC voltage and the slopes of the capacitor voltage are different for walking and running.}
	\label{fig:signal_comapre_activity}
\end{figure}

Because kinetic energy harvesting (KEH) methods harvest energy from external kinetic contexts (e.g., human motion, activities, mechanical vibrations) that strain or vibrate the transducer, the generated energy signal contains patterns and signatures of the external contexts. Thus, by using appropriate signal processing and pattern recognition algorithms, the KEH signal can be used as the proxy to sense and detect the external kinetic context. Following this intuition, researchers have attempted to detect a range of contexts directly from the KEH without using any conventional motion sensors (e.g., accelerometer). In the following, we give an overview of KEH-based sensing techniques for eight different contexts, discuss their performances as well as any reported power saving obtained when compared to the conventional sensor-based context sensing. 

\subsubsection{Human Activity Recognition (HAR)}
For KEH-powered wearable IoTs, such as fitness bands or smart shoes, the harvested energy is significantly influenced by the activity performed by the user. For two activities (e.g., walking and running), Figure~\ref{fig:signal_comapre_activity} illustrates both the AC voltage (instantaneous power) from the transducer as well as the capacitor voltage (accumulated energy) collected from a KEH-powered wearable worn by a test subject in our laboratory. In case of the instantaneous power signal, the maximum of the AC voltage signal generated by running is higher than that from walking. Similar behavior can be observed in Figure~\ref{fig:signal_comapre_activity}(b, in which the charging rate of the capacitor voltage (i.e., the slope of the signal) is distinct between the two activities. It means that human activity recognition (HAR) can be realized without engaging any specialized sensors. An important question is whether such KEH-based activity detection can scale beyond just walking and running. 

Using piezoelectric energy harvester inside shoes, the authors in~\cite{han2016self} conducted an experiment with three subjects performing six different activities -- normal walking, strolling, brisk walking, jogging, ascending stairs, and descending stairs. By simply using the relationships between peak values, time length, and slopes of the transducer AC waveforms, they were able to classify all six activities with over 90\% accuracy. This experiment demonstrates that it is feasible to use KEH as a self-powered sensor for HAR. 

In~\cite{khalifa2017harke}, the authors (that includes some of the authors of this paper) conducted further experiments to assess (a) how well the KEH can detect human activities when worn on other body parts, such as wrist, instead of in the foot\footnote{Note that it is relatively easier to detect human motions when the sensor is worn in the foot, but the performance drops when worn in hands.}, (b) how accurately the KEH can detect activities compared to the accelerometers, and (c) how much power can be saved if KEH, which does not require a power supply, is used for HAR instead of an accelerometer. The authors in~\cite{khalifa2017harke} also used piezoelectric KEH and conducted experiments with ten subjects performing five different activities -- standing, walking, running, ascending stairs, and descending stairs -- while holding the KEH device in the hand. To compare between the KEH and accelerometer, the handheld device was also instrumented with an accelerometer. Using machine learning, the authors found that the accelerometer could classify these five activities with an accuracy of 95\%, while KEH could achieve only 80\%. This outcome revealed that although KEH has the potential for HAR, it may be challenging to achieve recognition accuracy comparable to the conventional sensor-based systems, especially when the wearable is worn in the hand. On the positive side, a detailed power consumption experiment revealed that the KEH-based detection consumed 79\% less device power than the accelerometer-based system. In fact, the power consumption for KEH-based HAR can be saved even further when the accumulated energy in the capacitor is used once in a while to detect different activities instead of continuously sampling the transducer AC voltage to detect patterns~\cite{lan2017capsense}.

\subsubsection{Transportation Mode Detection}
As noted in Table~\ref{t:sources}, KEH can harvest energy from vibrations. Given that a wearable device is subjected to different vibration patterns when the user travels via different transportation modes, it is expected that KEH can be used to sense the transportation mode of an individual's everyday travel. Applying machine learning to KEH AC voltage time series data from a wearable piezoelectric energy harvesting device, Lan et~al.~\cite{lan2019entrans} was able to achieve 85\% accuracy in determining whether the user was traveling by car, bus, or train. A more detailed study further revealed that KEH data can also identify the specific train routes traveled by a user~\cite{abadi2016energy}.

\subsubsection{Estimation of Calorie Expenditure}
\begin{figure}[t]
\centering
\includegraphics[scale=0.58]{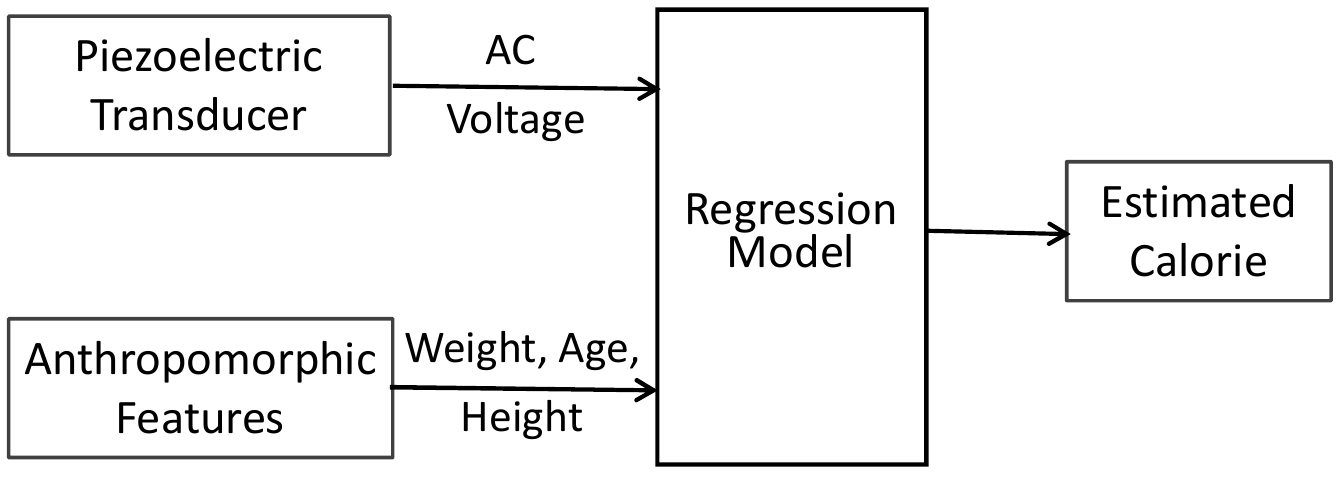}
\caption{Signal pipeline for calorie estimation from piezoelectric energy harvesting voltage.}
\label{f:calorie}
\end{figure}

During physical activities, people produce kinetic energy by expending or burning some calories. Since a wearable KEH harvests kinetic energy produced during physical activities, the voltage output of a wearable piezoelectric transducer should contain information that can be used to estimate the amount of calorie expended by the wearer. To prove this hypothesis, the authors in~\cite{lan2015estimating} conducted an experimental study with ten volunteers performing two different physical activities, such as walking and running. The expended calorie is estimated using the regression model $C = X \beta + \epsilon$ (the signal pipeline is shown in Figure \ref{f:calorie}), where C indicates the estimated calorie expenditure at the $k^{th}$ minute; and $X$ denotes the vector of input signals, including the anthropometric features of the volunteers (i.e., their weight, height, and age), as well as the output AC voltage signals from the transducer. The parameters $\beta$ and $\epsilon$ are the vector of coefficients and residual error, respectively. It was found that, for most subjects, the calorie estimations obtained from the output voltage of KEH were very close to those obtained from a 3-axial accelerometer for both walking and running.

\subsubsection{Step Counting}

Several efforts have been made on investigating the possibility of counting steps from the kinetic energy harvested from a wearable device. For example, in~\cite{rodriguez2017intermittently}, a ferroelectric energy harvesting device was designed in the form factor of an insole to harvest energy from human walking. The capacitor voltage waveform is leveraged for the purpose of step-counting, as it exhibits a stair-like pattern where each `stair' corresponds to a single step of walking. Experimental results showed an estimation error less than 4\% in step counting.

In~\cite{kalantarian2016pedometers}, the authors also used KEH inside shoes, but the capacitor voltage was used in a different way to count the steps. Once the capacitor voltage exceeds a pre-defined threshold, the capacitor would be discharged to power a Bluetooth beacon transmission, which is received by a nearby smartphone running the step counting app. When the capacitor voltage decreases below the voltage threshold, Bluetooth transmission would stop until the capacitor voltage reaches the threshold again. Thus, by counting the number of Bluetooth packets received, the smartphone app can estimate the number of steps taken. The reported error for step counting ranged from 4\% to 20\% for walking on a flat surface.

Finally, it was observed in~\cite{khalifa2015step} that the KEH power generation exhibits distinctive peaks for each step, which could be accurately detected using existing peak detection algorithms. They tested their peak detection based step counting algorithm with four subjects under different walking scenarios including stairs covering a total of 570 steps and were able to detect the steps with 96\% accuracy.

\begin{table*}
	\centering
	\caption{Summary of recent research on context sensing from energy harvesting signals.}
	\label{tab:sensing-application-overall}
	\ra{1.2}
	\begin{tabular}{clll}
\hline
		\textbf{Source}  & \textbf{Application} &\textbf{Method/Algorithm} & \textbf{Performance}      \\ \hline
		
		\multirow{17}{*}{Kinetic}  & \multirow{4}{*}{Activity recognition} & Waveform analysis~\cite{han2016self}  &  90\% accuracy on 6 activities with 3 subjects   \\ \cmidrule{3-4}
		
		&    & \multirow{2}{*}{Machine learning~\cite{khalifa2017harke,khalifa2015energy}}  &  80\% accuracy on 5 activities with 10 subjects, save 79\% power compared to  \\ 
		&    & &  accelerometer based method  \\ \cmidrule{2-4}
		
	   & Transport mode detection   & Machine learning\cite{lan2019entrans} &  85\% accuracy on 3 motorized modes    \\  \cmidrule{2-4}
	   
	   & Calorie estimation   &  Linear regression\cite{lan2015estimating}   &  Estimation accuracy close to that obtained from an accelerometer\\ \cmidrule{2-4}
	   
	    &  \multirow{4}{*}{Step counting}  &  Waveform analysis~\cite{rodriguez2017intermittently}    & Counting error less than 4\%\\ \cmidrule{3-4}
	    
	     &   &  Packet counting\cite{kalantarian2016pedometers}   &Counting error ranges from 4\% to 20\% on flat surfaces\\ \cmidrule{3-4}
	   
	   &    &  Peak detection\cite{khalifa2015step}   & 96\% accuracy with 570 steps under 4 surfaces including stairs\\ \cmidrule{2-4}
	   
	   &  Gait recognition  &  MSSRC\cite{dong2018sehs,xuTMCGait}   & 95\% accuracy with 20 subjects, save 92\% power compared to accelerometer\\ \cmidrule{2-4}
	   
	   &   Hotword detection  &  Machine learning\cite{khalifa2016feasibility}   & Up to 85\% accuracy when spoken from 3cm with 8 subjects \\ \cmidrule{2-4}
	   
	   &   Airflow speed monitoring  &  Peak detection \cite{li2018trinity}   & Estimation error of 0.2m/s \\ \cmidrule{2-4}
	   
	   &    Acoustic communication  &   ON-OFF keying \cite{lan2017veh,lan2018hidden}   & 5 bit/s data rate with a bit error rate of 1\% at distance of 80cm \\ \cmidrule{1-4}

	   	\multirow{4}{*}{Thermal} &Water flow detection& Binary indication~\cite{martin2012doubledip}  &  Extend the battery life of water flow sensors to 20 years   \\ \cmidrule{2-4}
	   		
	   	& Heat appliance monitoring& Packet interval~\cite{campbell2014energy}  & Successfully implemented to monitor stoves, radiators, and hot water flow  \\ \cmidrule{2-4}
	   	
	   	& Chemical reaction detection& Pulse amplitude~\cite{zarepour2015remote,zarepour2017semonnew}  & A remote receiver can detect the reaction type from the received
	   	pulse energy  \\ \cmidrule{1-4}

	   	\multirow{9}{*}{Solar} & Localization & Machine learning~\cite{randall2007luxtrace}  & Achieve a distance estimation accuracy of 21cm    \\ \cmidrule{2-4}
	   			
	   	&Positioning &Sunlight map~\cite{chen2016sunspot}  & Infer the longitude and latitude of a location   \\ \cmidrule{2-4}	
	   	
	   	& Place recognition & Machine learning~\cite{umetsu2019ehaas }  &86.2\% accuracy with 9 places using 5 types of solar cell   \\ \cmidrule{2-4}	
	   	
	   	& \multirow{3}{*}{Gesture recognition} & Peak counting\cite{varshney2017battery} &
	   	Detect 3 different hand gestures \\ \cmidrule{3-4}	
	   	
	   	&   & Machine learning\cite{ma2019solargest,ma2018gesture} &
	   	96\% accuracy with 6 gestures, save 44\% power compared to photodiode \\ \cmidrule{2-4}

	   & Visible light communication  & OFDM\cite{wang2015design} &11.84 Mbps data with a bit error rate of $1.6 \times 10^{-3}$ \\ \cmidrule{1-4}
	   
	   RF & Environment sensing&Impedance changing\cite{wang2018challenge} & Convert  RFID tags into light sensor and temperature sensor \\
		
		\hline
	\end{tabular}
\end{table*}

\subsubsection{Gait Recognition}

It is well known that human gait has distinctive motion patterns for different individuals, which can be detected accurately using accelerometers for user authentication~\cite{gafurov2006biometric,muaaz2017smartphone}. With 20 volunteers, the authors in~\cite{xuTMCGait} set out an experiment to determine whether the KEH AC voltage signal can also recognize human gait. They considered two different types of KEH, one based on piezoelectric energy harvester and the other on electromagnetic energy harvester. The authors found that both types of transducers were able to detect gait, but with conventional classification techniques, which operate
over a single step, the KEH-Gait achieves approximately 6\% lower accuracy compared to the accelerometer-based gait recognition. They proposed a novel classification method, called Multi-Step Sparse Representation Classification (MSSRC), which can match the performance to that of accelerometer by intelligently fusing information from multiple steps. They also showed that by eliminating accelerometer from the processing pipeline, the KEH-based gait recognition can reduce power consumption by 92\%.

The experiments in~\cite{xuTMCGait} employed isolated transducers without connecting them to a capacitor to store energy. Later, it was discovered in~\cite{dong2018sehs} that when a capacitor is used to store energy, the capacitor voltage interferes with the AC voltage waveform of the transducer, which reduces gait recognition accuracy. To address this problem, the authors in~\cite{dong2018sehs} proposed a filter to minimize the influence of the capacitor voltage on the instantaneous AC signal, thereby achieving simultaneous energy harvesting and sensing for KEH. 

\subsubsection{Hotword Detection}

Human voice creates vibrations in the air, which could potentially be picked up by the KEH hardware inside a mobile device. Based on this observation, in ~\cite{khalifa2016feasibility} the KEH's feasibility and accuracy were studied for detecting \textit{hotwords}, such as ``OK Google", which are used by voice control applications to delineate user commands from background conversations. Using eight subjects,  the authors evaluated two types of hotword detection, susch as speaker-independent (which does not
require speaker-specific training) and speaker-dependent (which relies on speaker-specific training). They found that when spoken from 3cm, piezoelectric transducers can detect hotwords with accuracies of 73\% and 85\%, respectively, for speaker-independent and speaker-dependent detection. They also reported that the orientation of the piezoelectric beam relative to the speaker has a significant impact on the hotword detection accuracy.

\subsubsection{HVAC Airflow Monitoring}

The IoTs can play a significant role in reducing energy consumption of heating, ventilation and air conditioning (HVAC) systems by adaptively controlling its output (e.g., air speed through the vents) to the current population density of the serving area. To this end, small IoT devices fitted with sensors have to be installed near the vents to periodically measure the output air speed, temperature, etc. and send them to a smart home control box. For these IoTs to be self-powered with energy harvesting, the sensing power consumption has to be minimal. However, typical airflow sensors consume on the order of hundreds of mWatts, which was a motivation in~\cite{li2018trinity} to study the feasibility of HVAC airflow sensing using the voltage of piezoelectric energy harvesters. Through experiments, the authors in~\cite{li2018trinity} demonstrated that the peak voltage of the piezoelectric harvester is a function of the airflow speed and the voltage value under a given speed has a variation only up to 0.06V, which suggests that using voltage to infer airflow speed has an error of only 0.2m/s. They were able to `catch' the peak voltage by sampling the transducer output only once every 100 ms, resulting in an overall power consumption of only 500 $\mu$W when KEH was used as an airflow sensor.

For on-demand infrastructural sensing (i.e., only sense the context when desired events occur), conventional approaches usually adopt duty-cycling or adaptive sampling to save the system power. However, using energy harvester as a passive event detector further pushes the power consumption to zero. In~\cite{liu2019ecovibe}, the authors presented ECOVIBE, which exploits a vibration energy harvester to detect the arrival of trains and then activates the sensors for rail condition monitoring. Meanwhile, the harvested energy was used to power the sensors during the monitoring process. 

\subsubsection{Acoustic Communication}

In~\cite{lan2017veh} the piezoelectric KEH was leveraged as a communications receiver to receive data packets transmitted by a nearby speaker by modulating sound waves within the resonance frequency of the transducer using simple ON-OFF keying. Experiments with a real KEH device revealed that, at a distance of 80 cm, a laptop speaker with the proposed modulation scheme can successfully transmit information to the KEH device at 5 bps at a target bit error rate of less than 1\%. While this scheme would enable a KEH-powered IoT to receive commands from other speaker-enabled devices, such as a laptop, TV, music player, etc., acoustic data transmissions would be audible and could be annoying to some users. A more advanced modulation scheme~\cite{lan2018hidden} was later proposed that enables the transmitter to completely hide the acoustic data transmission within background music so the data transmission is not audible anymore.

\subsection{Context Sensing from Thermoelectric EH}

As discussed in Section \ref{section:iots}, thermoelectric energy harvesting can convert any temperature difference in space or time into usable electricity. Therefore, it is possible to use a thermoelectric energy harvester as a sensor to detect temperature-related contexts. Indeed, researchers have successfully demonstrated such potentials for detecting water flow, activities of various heating appliances, as well as chemical reactions in a reactor.

\subsubsection{Water Flow Detection}
Currently, wireless water flow monitoring in residential water fixtures requires installing sensors with access to electrical wiring or replacing batteries frequently. In~\cite{martin2012doubledip} was proposed a thermoelectric energy harvesting solution, called DoubleDip, which harvests energy from the pipe's thermal gradient, i.e., the temperature difference between the pipe and the room temperature, when hot water flows through the pipe. The harvested thermal energy is used to wake up the sensor from deep sleep mode as well as to compensate battery energy expenditure. Because the sensor automatically wakes up only when needed, instead of duty cycling at fixed intervals, it can save energy significantly. The authors claimed that DoubleDip can extend the battery life of water flow sensors to 20 years or even to perpetuity. 

\subsubsection{Heat Appliance Monitoring}
The authors in~\cite{campbell2014energy} proposed and demonstrated that thermoelectric generators (TEGs) can be effectively used as sensors to remotely monitor any heat-generating appliances. The key idea is to wake up the sensing device when sufficient energy has been harvested and force it to transmit small wireless packets as long as there is energy available to harvest, which is basically the same principle applied to~\cite{kalantarian2016pedometers} to sense human steps using KEH. Since thermal energy is produced only when the appliance becomes active, it is then possible to monitor the activity of the appliance by simply monitoring the timing of the received packets at a nearby receiver. The authors in~\cite{campbell2014energy} implemented this idea to monitor stoves, toaster ovens, radiators, and hot water flow through shower heads.  

\subsubsection{Chemical Reaction Detection}

In the context of Internet of nano things, a self-powered sensing architecture, called SEMON, was proposed in~\cite{zarepour2015remote,zarepour2017semonnew} for remote detection of chemical reactions. In particular, pyroelectric nanogenerators that can harvest energy from the temperature variation in time domain, are fitted with Graphene-based nano-antennas radiating in the Terahertz band (0.1-10THz) and embedded in the catalyst surface where different types of chemical reactions take place. Each reaction consumes or dissipates some heat, which causes temperature fluctuations on the catalyst surface. When a particular reaction takes place on the catalyst surface, a SEMON node harvests the energy consumed or dissipated by the reaction and turns it into a Terahertz radio pulse. Specifically, a pyroelectric nanogenerator harvests electrical energy from each temperature fluctuation and uses the energy to transmit a THz pulse with an amplitude proportional to the harvested energy. Because different types of reactions dissipate different amounts of energy, the authors were able to show via simulation experiments that a remote receiver can detect the reaction type from the received pulse energy.

\subsection{Context Sensing from Solar EH}

Similar to KEH revealing motion patterns or thermoelectric energy harvester recognizing temperature changes, the solar energy harvesting provides information about lighting conditions. A wide range of light-based sensing applications, including localization, gesture recognition, and even data transmission, have been proposed based on the characteristics of solar energy harvesting. We survey these developments in this section.

\subsubsection{Localization and Positioning}

Because different locations experience different lighting conditions under the same lighting infrastructure, it is possible to detect location by analyzing the received lights. In~\cite{randall2007luxtrace}, the authors proposed and prototyped a wearable solar cell based indoor positioning system called LuxTrace. In their prototype, solar cells are attached to the shoulder of the user, which not only harvest energy from the ambient indoor light, but also detect the received light strength (RLS). By utilizing the RLS, a trained model was exploited to estimate the relative distance between the user's location and the light sources. Experimental evaluation indicated that LuxTrace can achieve a distance estimation accuracy of 21 cm (80\% quantile).

In~\cite{umetsu2019ehaas}, different types of solar cells are exploited to identify places in daily living spaces. The underlying principles are: (1) the amount of energy generated by solar cells is almost linear with the environment illuminance, and (2) solar cells manufactured with different materials, e.g., silicon and organic, show distinct response to the wavelength of light. Using machine learning with the data collected from five different types of solar cells in nine different places, such as laboratory, toilet, elevator, outdoor, etc., the authors demonstrated that a place recognition accuracy of 86.2\% can be achieved using two types of solar cells.  

The idea of using light to detect location can be extended to a global scale. Every location on Earth has a unique solar signature, like a unique sunrise and sunset time, which can be used to create a \textit{sunlight map} of the Earth. Such solar signatures were used in~\cite{chen2016sunspot} to design a system, called SunSpot, which is able to infer a location's longitude and latitude separately, based on the sunlight map.

\subsubsection{Gesture Recognition}

Many IoT devices in the future are likely to use solar panels for energy harvesting. By moving a hand close to the solar panel, it is possible to influence the solar energy harvesting, which in turn can be exploited to realize gesture recognition capabilities in any solar-powered IoT device. Motivated by this idea, gesture recognition capability of different types of solar cells have been investigated in different conditions. For conventional opaque solar cells, a thresholding circuit was designed in~\cite{varshney2017battery} to detect if the solar energy falls below a given threshold. This allows to detect when a hand is moved very close to the solar panel, which causes the solar energy to drop below the threshold. By simply counting the number of times the hand is moved near the solar panel, the authors were able to detect three different hand gestures.

\begin{figure}[]
	\centering
	\centering
		\includegraphics[scale=0.52]{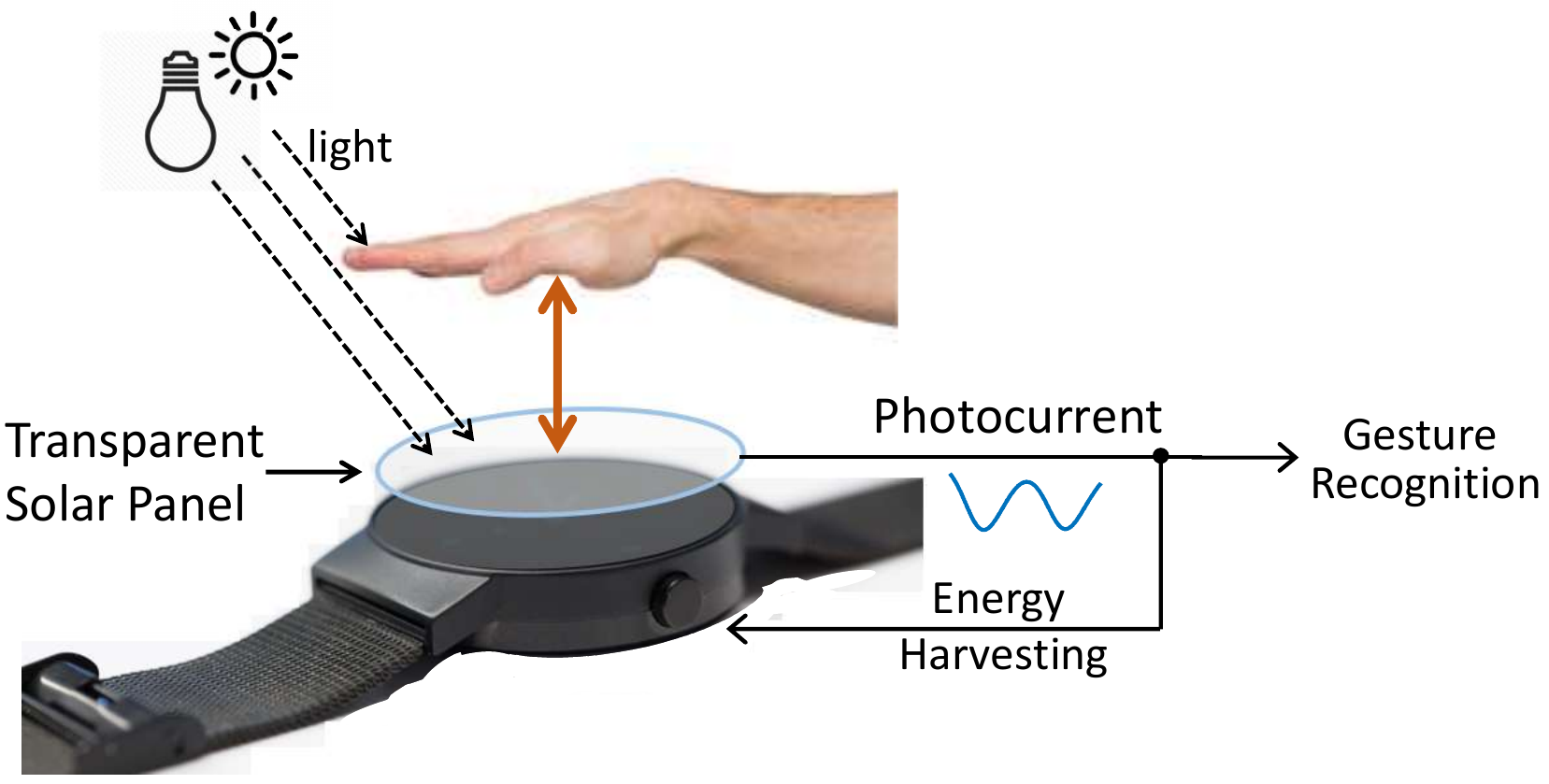}
	\caption{Illustration of a transparent solar powered smartwatch with solar-based gesture recognition.}
	\label{fig:solarwatch}
\end{figure}

\begin{figure}[]
	\centering
	\centering
		\includegraphics[scale=0.62]{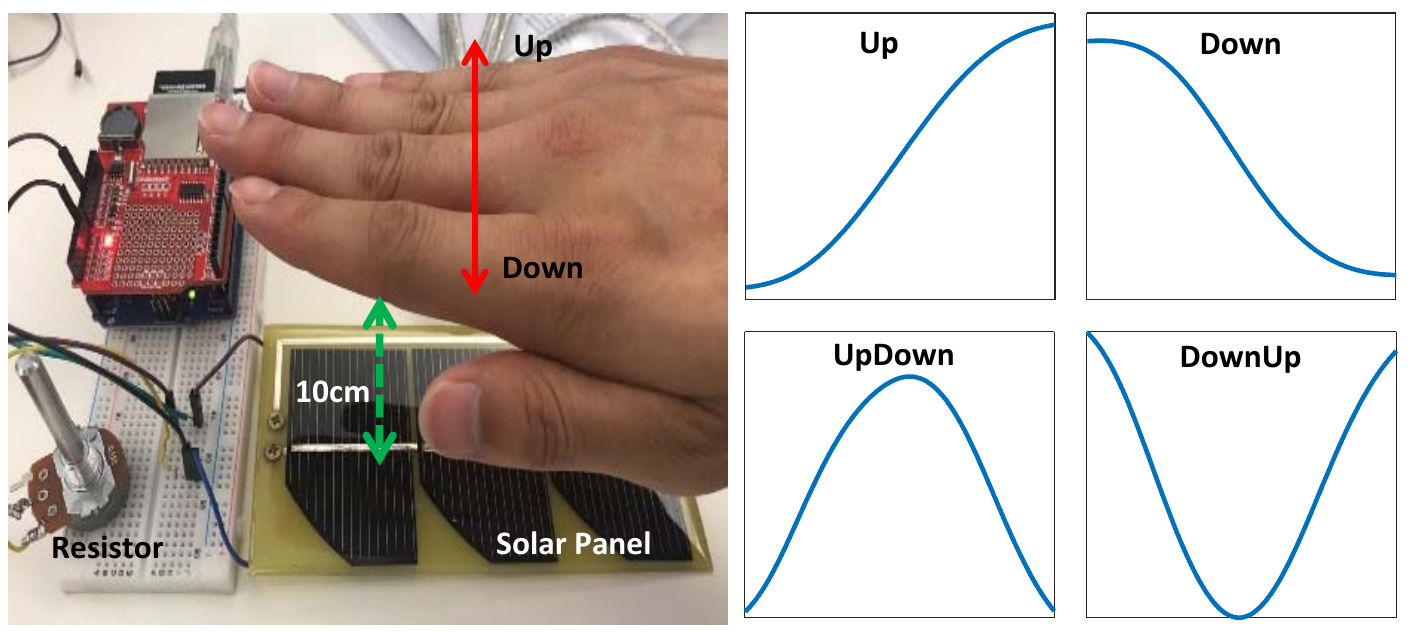}
	\caption{Illustration of the potential for solar energy harvesting to detect gestures. The left figure shows the experiment setup and the right shows the distinct solar photocurrent waveforms generated under four different hand gestures.}
	\label{fig:SolarExpeiriment}
\end{figure}

The gesture recognition feasibility was investigated in~\cite{ma2018gesture} for \textit{transparent} solar cells, an emerging solar energy harvesting technology that allows us to see through those cells. This revolutionary discovery is creating unique opportunities to turn any mobile device screen, such as that of a smartwatch (see Figure \ref{fig:solarwatch}), into solar energy harvester. Transparency, however, means that the absorption efficiency of the solar cell in the visible light band is significantly lower compared to the opaque cells. The lower absorption rate results in weaker responsiveness to the visible light, thus making gesture recognition challenging. To overcome this challenge, machine learning was considered to analyze the solar photocurrent waveforms produced during different gestures. As shown in Figure~\ref{fig:SolarExpeiriment}, the waveforms of different gestures are different, which can be detected using machine learning. Experimental results from~\cite{ma2018gesture} demonstrate that five hand gestures can be detected by transparent solar cells with an average accuracy of 95\% and transparent solar cells can recognize some of these gestures almost as good as the opaque cells.

A key challenge for solar-based gesture recognition research is the lack of access to many new types of solar cells, such as transparent solar cells, which still remain in research labs. This makes solar energy harvesting based sensing research out of reach. To facilitate gesture recognition research with solar cells, the authors in~\cite{ma2019solargest} developed a simulator, called SolarGest, which can be used to generate photocurrent waveforms for any arbitrary solar cells and hand gestures (the simulator code is released for public use~\cite{simulator}). They validated the simulator through various gesture experiments with both opaque and transparent solar cells. To further improve the robustness of solar-based gesture recognition under non-deterministic operating conditions, they combined dynamic time warping with Z-score transformation in a signal processing pipeline to pre-process each gesture waveform before it was analyzed for classification. Their experiments with 6,960 gesture samples for six different gestures revealed that even with transparent cells, SolarGest can detect 96\% of the gestures while consuming 44\% less power compared to the light sensor based systems.

\subsubsection{Visible Light Communication}

Visible light communications (VLC) is seen as a promising new alternative to the conventional RF-based data communication. The VLC harnesses a portion of the
electromagnetic spectrum that is currently license-exempt and offers a vast amount of bandwidth for high-speed wireless data communications without any 
interference to existing radio communication systems. The VLC uses light emitting diode (LED) as transmitter and photodiode as receiver. However, these photodiodes require a power supply to operate. In~\cite{wang2015design} is proposed the use of standard solar panels as VLC receivers that can demodulate VLC data signal without the need of an external power supply. Using orthogonal frequency division multiplexing (OFDM), the authors were able to achieve a data rate of 11.84 Mbps with a bit error rate (BER) of $1.6 \times 10^{-3}$. This outcome suggests that simultaneous communication and energy harvesting can be realized with solar panels.

\begin{figure*}[]
\centering
\subfigure[]{
\includegraphics[scale=0.7]{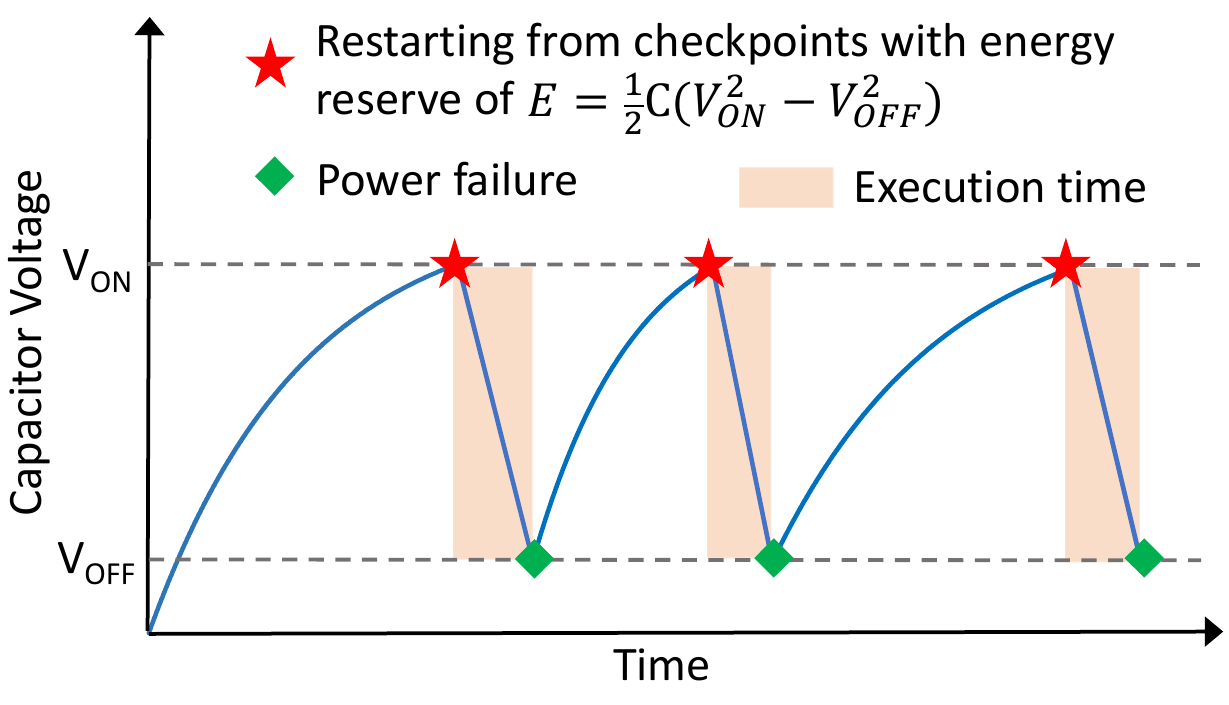}}
\subfigure[]{
\includegraphics[scale=0.7]{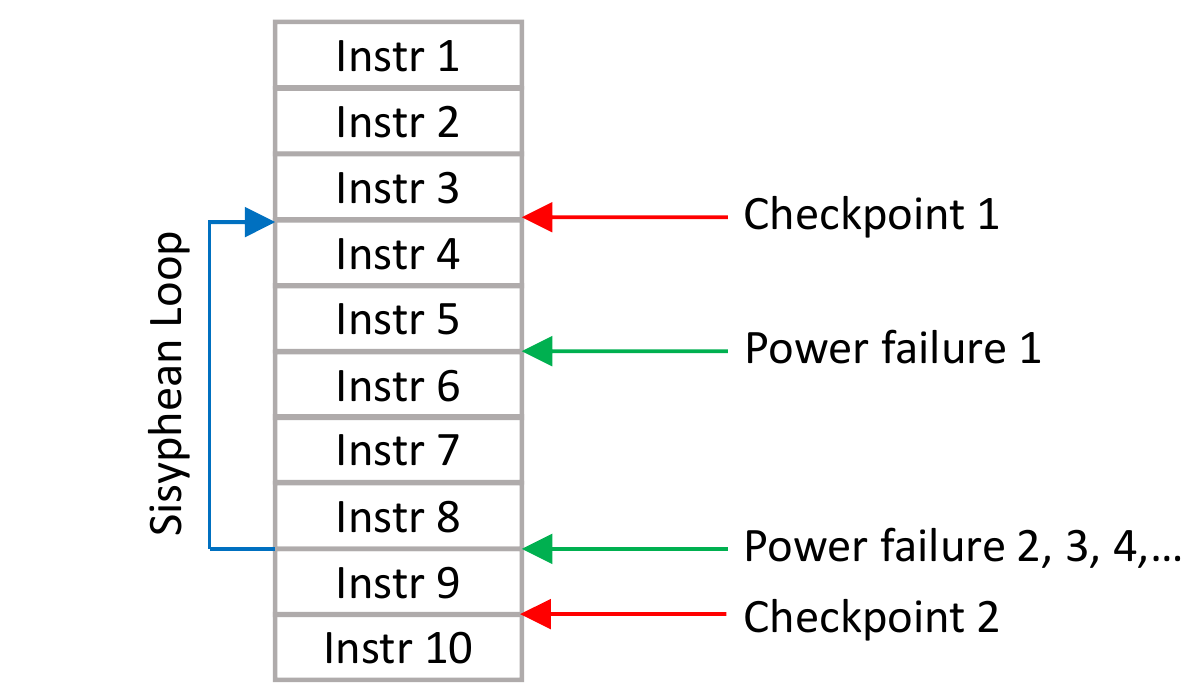}}
\caption{Illustration of (a) intermittent power supply and execution. (b) Sisyphean execution.}
\label{f:sisy}
\end{figure*}

\subsection{Context Sensing from RF EH}

Passive radio frequency identification (RFID) tags are batteryless devices ubiquitously deployed today to monitor different types of events. These tags are powered by inductive coupling with RF energy generated by a nearby transmitter known as RFID reader. An RFID tag has an antenna which harvests RF energy from the reader's transmission and also reflects the reader's signal to encode the ID of the tag. With conventional RFID, the reader is only able to read the ID of the tag. These tags are often embedded in animals and objects that are being monitored. If a reader can detect the presence of a given ID at certain location and time, then the knowledge can be used to monitor the activity or trajectory of the animal or object.

In~\cite{wang2018challenge}, embedding passive sensors, such as phototransistors whose resistance depends on the amount of ambient light received, was explored into RFID tags to alter the impedance of the antenna and hence the amount of RF energy radiated from its antenna. This would make the received signal strength (RSS) at the reader dependent on the current lighting condition of the RFID tag. The authors of~\cite{wang2018challenge} have demonstrated this outcome for both light sensing (using phototransistor) and temperature sensing (using thermistor) using batteryless RFID tags. The authors claim that the idea is generic and can be used to sense a range of other contexts including detecting color, humidity, pressure, or even hand gestures.

\subsection{Lessons Learned}

Our survey shows that energy harvesters can definitely work as a virtual sensor to detect a wide variety of contexts. Below we highlight some important lessons learned from this study. 

\begin{itemize}
\item Out of the four typical EH sources, kinetic energy harvesting has been explored the most for context detection, which suggests that EH-based context sensing is most promising to replace conventional motion sensors, such as accelerometers and gyroscopes, in active scenarios. Because temperature usually changes gradually, thermal EH is rather less sensitive to the context. Although Solar and RF EH based context sensing is still in their infancy, there is a great potential to develop novel applications in the future, as light and radio waves are ubiquitous and change with context rapidly and dramatically.

\item Reusing energy harvesters as sensors can reduce the EH-IoT system power consumption significantly, thereby prolonging the device operation time.

\item While more significant power saving could be achieved for context detection from the accumulated energy signal (detecting user activity by reading the capacitor charge level), its detection accuracy is expected to be lower than that of instantaneous EH signals (e.g., AC voltage signal from a piezo transducer).

\item Unlike specialized sensors that are designed to produce very accurate measurements of physical phenomena, energy harvesters can only act as crude sensing hardware. Consequently, recent studies have found that their context detection performance is inferior to that of specialized sensors. Clearly, further research and development are required before energy harvesters could replace specialized sensors.
\end{itemize}

\section{Intermittent Computing for Batteryless EH-IoTs}
\label{section:computing}

Unlike batteries, power supply from energy harvesting is uncertain and often unpredictable. As a result, it is not unlikely for an EH-IoT device to experience frequent power outages, which can be as extreme as several times per second for certain devices~\cite{WISP}. As all variables and registers stored in the volatile memory during program execution are completely lost when the power supply fails, the program execution on EH-IoTs must rely on \textit{checkpointing}, a technique to periodically save volatile states in non-volatile memory, so that when the power returns, the program execution can restart from a known state. However, the unstable energy supply poses unique challenges on the operation of EH-IoTs and conventional checkpointing is unable or inefficient for intermittent computing. In this section, we identify the challenges encountered for intermittent computing of EH-IoTs, followed by the review existing efforts on addressing such challenges.

\begin{figure*}[]
	\centering
	\centering
		\includegraphics[scale=0.7]{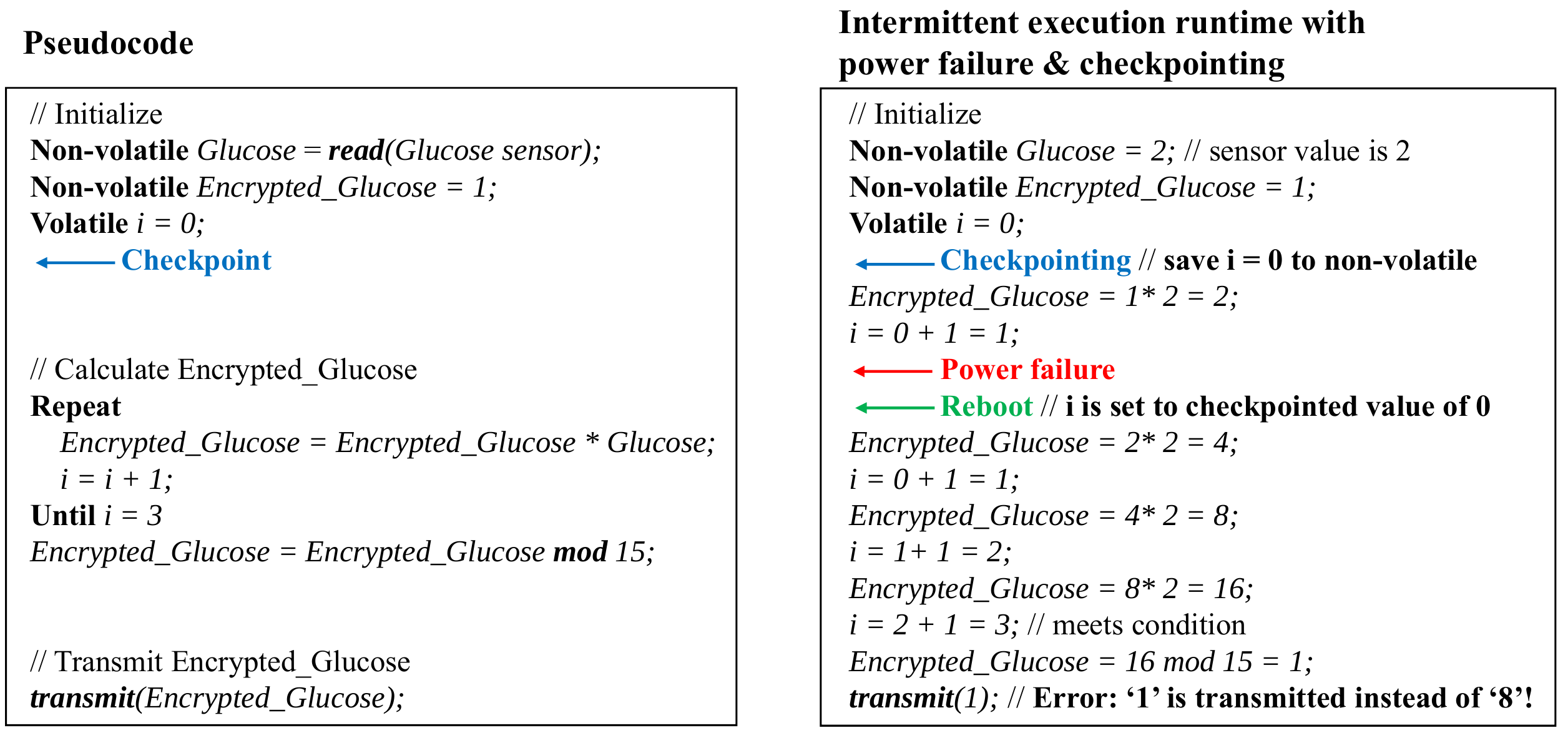}
	\caption[Caption for LOF]{An example of encrypting glucose reading using RSA public key (3,15)\footnotemark. The original glucose reading, stored in variable Glucose, represents one of the 15 different levels of glucose denoted as 0 to 14. The encrypted value is obtained as $Glucose^{3} \ mod \ 15$. (a) The source pseudocode showing that all variables except index $i$ are non-volatile. (b) Execution of the code with a glucose reading of 2, encrypted as $2^3 \ mod \ 15 = 8$. With the power failing after completing the loop once, the index $i$ is reset to 0 and the loop is eventually executed 4 times producing the encrypted value of 1, instead of 8.}
	\label{fig:data-consistency}
\end{figure*}
\footnotetext{The key is obtained by $p=3, q=5 \Rightarrow n=3 \times 5=15, E=3$.}

\subsection{Challenges of Intermittent Computing}
Although checkpointing can be effective to survive power failures, it gives rise to the following performance issues for an intermittently powered EH-IoT~\cite{lucia2017intermittent}:
\subsubsection{Energy Overhead of Checkpointing}
During runtime, the program has to copy all state variables from volatile memory to non-volatile memory at each checkpoint, which will then have to be reloaded to volatile memory when power returns. Thus, the total energy overhead of checkpointing can be significant if checkpointing is used frequently. In addition, the number of state variables to be saved vary when the checkpointing at different stages of a program. For example, less volatile variables need to be copied to non-volatile memory if the checkpointing process is executed right after a function call rather than during the function call. Therefore, proper selection of the stage/location for checkpointing is critical. 

\subsubsection{Sisyphean Execution} 
As mentioned earlier, checkpointing allows a device to roll-back on the execution time-line and start from a previously saved known state when power returns following an earlier failure. Thus, at minimum, power failures will cause some roll-back preventing smooth execution of program on EH-IoT devices. A more serious problem can occur when the accumulated energy since the previous power cut-off is not enough to reach the next checkpoint. In that case, the program makes a small progress each time it is powered on only to roll-back to the same checkpoint repeatedly, making the programming task impossible to complete\footnote{This problem is referred to as Sisyphean because in Greek mythology, Sisyphus, a king who annoyed the gods, was condemned for eternity to roll a huge rock up a steep hill, only to watch it roll back down.}. 

Figure \ref{f:sisy} (a) illustrates intermittent power supply and execution. Power fails and program execution stops when capacitor voltage falls below a minimum threshold ($V_{OFF}$). Power comes back and execution restarts from the last checkpoint when capacitor voltage reaches to $V_{ON}$. Figure \ref{f:sisy} (b) presents an example of Sisyphean execution. Specifically, the first power failure occurs after completing the 5th instruction; the execution restarts from instruction 4 (checkpoint 1), but the energy reserve runs out just before reaching checkpoint 2. Thus, the execution reaches up to instruction 8 only to roll back to instruction 4 repeatedly forming a `Sisyphean loop'.

\subsubsection{State inconsistency}  
Checkpointing can guarantee 100\% state preservation only if all variables are exclusively maintained either in volatile memory or in non-volatile memory, but not split across both of them. When all variables are maintained in volatile memory, they all are saved in non-volatile memory at the same time at a given checkpoint, and hence the system can roll back to a consistent global state after power failure. This is also the same when all variables are maintained in non-volatile memory. However, neither is practical because volatile memory is fast but expensive, but non-volatile memory is slow and less expensive. As a result, to achieve fast execution at a reasonable cost, most commodity hardware uses a small volatile memory to save the most frequently used variables and stores the rest in non-volatile memory. Thus the state of a program is split over volatile and non-volatile memories. When the system rolls back to the previous checkpoint, all variables in volatile memory rolls back to their previous values, but those in the non-volatile memory preserve their new values. As a result, data inconsistencies can arise in certain cases as illustrated in the RSA encryption example in Figure~\ref{fig:data-consistency}.       

\subsubsection{Timing Inconsistency}  
A wearable IoT may be tasked with monitoring a range of physiological data, such as blood pressure, heart rate, glucose level, and so on, and reporting them with timestamps to a cloud-based analytic for further processing.  However, if there is a power failure after collecting data from a limited set of sensors, the device can only read the remaining sensors when the power comes back. In this case, the timestamps for different set of sensors should be different, which could be reported accurately only if the device had persistent timekeeping across power failures. In battery powered devices, a real-time clock (RTK) is always maintained to avoid this problem. However, the RTK requires a power supply and hence persistent RTK cannot be supported in EH-IoTs which experience frequent power failure~\cite{hester2016persistent}. As a result, the device may report inconsistent time stamps for the sensor data.

To address the aforementioned issues, many strategies have been proposed to optimize checkpointing for batteryless EH-IoTs as well as solutions to maintain timing across power failures (see Table~\ref{tab:checkpointing} for a summary).

\begin{table*}[t]
\centering
\caption{Summary of recent research on computing optimizations for batteryless EH-IoTs.}
\label{tab:checkpointing}
\setlength\tabcolsep{2.2pt}
\ra{1.2}
\begin{tabular}{@{}llcccc@{}}\toprule

 \textbf{Literature} & \textbf{Technique}  &\textbf{Checkpoint overhead} &\textbf{Sisyphean execution} &\textbf{State inconsistency}  & \textbf{Time inconsistency}\\ \midrule

\multirow{2}{*}{Mementos~\cite{ransford2012mementos}} &CFG-based checkpoint placement, &  \multirow{2}{*}{\checkmark}  & \multirow{2}{*}{\checkmark}  &  \multirow{2}{*}{$\times$} &  \multirow{2}{*}{$\times$}   \\
  &voltage polling checkpoint activation &   & & &    \\ \cmidrule{1-6}
  
\multirow{2}{*}{Idetic\cite{mirhoseini2013idetic}} &CDFG-based checkpoint placement, &  \multirow{2}{*}{\checkmark}  & \multirow{2}{*}{\checkmark}  &  \multirow{2}{*}{$\times$} &  \multirow{2}{*}{$\times$}   \\
    &voltage polling checkpoint activation &   & & &    \\ \cmidrule{1-6}
 
QUICKRECALL\cite{jayakumar2014quickrecall} &Run time checkpointing and &  \multirow{2}{*}{$\times$}  & \multirow{2}{*}{\checkmark}  &  \multirow{2}{*}{\checkmark} &  \multirow{2}{*}{$\times$}   \\
  Hibernus\cite{balsamo2015hibernus,balsamo2016hibernus++}   &sleeping, advanced ADC interrupt
   &   & & &    \\ \cmidrule{1-6}

\multirow{2}{*}{DINO \cite{lucia2015simpler}} &Task-based checkpointing-versioning, &  \multirow{2}{*}{\checkmark}  & \multirow{2}{*}{$\times$}  &  \multirow{2}{*}{\checkmark} &  \multirow{2}{*}{$\times$}   \\
&manual idempotency analysis&   & & &    \\ \cmidrule{1-6}

\multirow{2}{*}{Ratchet\cite{van2016intermittent}} &Task-based checkpointing, &  \multirow{2}{*}{\checkmark}  & \multirow{2}{*}{$\times$}  &  \multirow{2}{*}{\checkmark} &  \multirow{2}{*}{$\times$}   \\
&automatic idempotency analysis&   & & &    \\ \cmidrule{1-6}

\multirow{2}{*}{Chain\cite{colin2016chain}} &Idempotency task programming, &  \multirow{2}{*}{\checkmark}  & \multirow{2}{*}{\checkmark}  &  \multirow{2}{*}{\checkmark} &  \multirow{2}{*}{$\times$}   \\
&channel-based data exchange&   & & &    \\ \cmidrule{1-6}

\multirow{2}{*}{HarvOS\cite{bhatti2017harvos}} &CFG-based checkpoint placement, &  \multirow{2}{*}{\checkmark}  & \multirow{2}{*}{$\times$}  &  \multirow{2}{*}{$\times$} &  \multirow{2}{*}{$\times$}   \\
&advanced ADC interrupt&   & & &    \\ \cmidrule{1-6}

\multirow{2}{*}{ Clank\cite{hicks2017clank}} &Dynamic idempotency task decomposition, &  \multirow{2}{*}{\checkmark}  & \multirow{2}{*}{\checkmark}  &  \multirow{2}{*}{\checkmark} &  \multirow{2}{*}{$\times$}   \\
&checkpointing and versioning&   & & &    \\ \cmidrule{1-6}

\multirow{2}{*}{Alpaca\cite{maeng2017alpaca}} &Idempotency task programming, &  \multirow{2}{*}{\checkmark}  & \multirow{2}{*}{\checkmark}  &  \multirow{2}{*}{\checkmark} &  \multirow{2}{*}{$\times$}   \\
&privatization data exchange&   & & &    \\ \cmidrule{1-6}

\multirow{2}{*}{Mayfly\cite{hester2017timely}} &Idempotency task programming, &  \multirow{2}{*}{\checkmark}  & \multirow{2}{*}{\checkmark}  &  \multirow{2}{*}{\checkmark} &  \multirow{2}{*}{\checkmark}   \\
&remanence timekeeping&   & & &    \\ \cmidrule{1-6}

CleanCut\cite{colin2018termination} &Automatic idempotency task decomposition &  \checkmark & \checkmark  & \checkmark &  $\times$  \\ \cmidrule{1-6}

TARDIS~\cite{rahmati2012tardis} &SRAM decay timekeeping &  NA & NA  & NA &  \checkmark  \\ \cmidrule{1-6}

CusTARD~\cite{hester2016persistent} &Capacitor voltage decay timekeeping &  NA & NA  & NA &  \checkmark  \\ 

 \bottomrule 
\multicolumn{6}{l}{`\checkmark' represents the corresponding challenge was addressed in this work, while `$\times$' means not. `NA' means not applicable.}
\end{tabular}
\end{table*}

\subsection{Checkpointing Optimizations}

\subsubsection{Checkpoint Placement and Activation} refers to the scheme of inserting potential checkpoints to the program at compile-time and activate checkpointing processes at run-time. Inserting checkpoints at different locations of the program leads to distinct checkpointing overhead (including energy and memory) as the number of variables to be saved varies. In addition, checkpointing too early before a power failure results in a waste of energy that can be used to perform more computations; similarly the checkpointing process cannot be completed if it is too late as the residual energy is not enough. Thus, the checkpoint placement should take into consideration the checkpointing overhead as well as device residual energy. Due to unpredictable power failures, however, optimal reasoning about the potential checkpoints is extremely difficult and imposing a heavy burden on the programmer.

Momentos~\cite{ransford2012mementos} is the first work that supports the execution of long-running programs on intermittently powered devices, using checkpointing approach. It required the programmer to manually place trigger points (i.e., potential checkpoints) in the code. The authors proposed three checkpoint placement strategies based on the Control Flow Graph (CFG), which is a graphical representation of a program including basic constructs (e.g., branching statements, loops, etc.) and the edges connecting these constructs. Specifically, Mementos placed a trigger point at each loop latch (\textit{loop-latch mode}) or after each call instruction (\textit{function-return mode}), which allows an energy check after each loop iteration or at each time a function returns, respectively. In the \textit{timer-aided mode}, a timer interrupt is triggered to periodically measure the supply voltage and activate the checkpoint, if necessary. The underlying placement rationale is that the loop ends and function returns are locations where one may expect the stack to store fewer data, thereby less checkpointing overhead. Similarly, HarvOS~\cite{bhatti2017harvos} located lightweight checkpoints based on the information provided by the CFG of a program. The minor difference is that HarvOS also exploited the memory allocation pattern derived from static code analysis techniques~\cite{hofmann2003static} to accurately calculate the amount of memory allocated during the program.

However, placing checkpoints purely based on CFG may not be energy efficient in some cases. For instance, when power loss happens right before a function return, the program will restart from the previous checkpoint that is located before the function starts, therefore wasting the energy to re-execute the whole function. To address this problem, Idetic~\cite{mirhoseini2013idetic} was developed to optimally insert checkpoints, considering not only the checkpointing overhead but also the re-computation energy cost. This is achieved by leveraging the information provided by the Control Data Flow Graph (CDFG), an intermediate representation of a program that lies between the high-level behavioral specifications and the low-level Hardware Description Language (HDL), which models the connections and dependencies between processes. Different from CFG that only represents the control flow of a program, CDFG can present the data flow between different constructs as well. Idetic then formulated an optimization problem and derived optimal checkpoints using dynamic programming.

Actually, it is not necessary to initiate a checkpointing process at each checkpoint inserted in advance. For example, if the residual energy is enough to support the computations until next checkpoint, then checkpointing of the current system state is a waste of energy. Thus, energy-aware checkpoint activation at run-time was proposed~\cite{ransford2012mementos,mirhoseini2013idetic,jayakumar2014quickrecall}. At each checkpoint, currently available energy of the device is estimated by measuring the storage capacitor voltage and calculating the equation $E=1/2CV^2$, where $C$ is the capacitance of the capacitor and $V$ is the measured voltage. Then, a threshold based strategy is used to determine whether to activate a checkpoint. In detail, if the available energy is higher than a pre-defined threshold, the current checkpoint would be skipped and the execution continues; otherwise, the current system state would be checkpointed.

However, since the number of computations between every two successive checkpoints may be different, the determination of the pre-defined threshold is challenging. Usually, the threshold is derived through complex offline emulation~\cite{ransford2012mementos,balsamo2015hibernus}. In HarvOS~\cite{bhatti2017harvos}, the decision on whether to activate a checkpoint depends not only on the currently available energy but also on the worst-case estimation of the energy required to reach the next checkpoints, where the worst-case assumes that no energy will be harvested before the next checkpoint. This strategy ensures that the checkpoints are activated much closer to the last practical point where the system should take a checkpointing process, thereby reducing the energy waste on uncheckpointed work and unnecessary checkpointing.

\subsubsection{Checkpointing at Run Time} 
Some works like QUICKRECALL~\cite{jayakumar2014quickrecall}, Hibernus~\cite{balsamo2015hibernus}, and Hibernus++~\cite{balsamo2016hibernus++} leveraged the concept of checkpointing to ensure progress of a long-running program, but does not require insertion of checkpoints in advance. The underlying idea is that it initiates a checkpointing process only when a power failure is imminent, and then \textit{sleeps immediately}. Otherwise, it continuously executes programs without interruption. Similarly to the basic checkpointing method, the decision to take a checkpointing (or the judgment that a power failure is about to happen) is based on the measurement of currently available energy and the threshold based activation. As there is no pre-inserted checkpoint, continuously polling the capacitor voltage is needed, which incurs heavy energy consumption. As a result, Hibernus and Hibernus++ exploited an advanced analog-to-digital converter (ADC) that has interrupt functionality to detect whether the voltage reaches a certain threshold, thereby significantly reducing the power consumption of voltage polling. However, how to determine the threshold still remained a challenge and offline emulation was conducted in Hibernus. To avoid the risk of power failure when saving the system states, Hibernus++ implemented a decoupling capacitor that specially provides power supply to the device during the checkpointing process. 

Run-time checkpointing eliminates the burden on the programmers to reason and insert potential checkpoints at compile-time. Moreover, since it sleeps immediately after checkpointing, data inconsistency issue does not exist. However, two drawbacks arise. First, it assumes that the closer the checkpointing to a power failure, the better the program executes, which may not be the fact as it ignores the energy and memory overhead of checkpointing. For example, the last moment before power loss may have a large amount of volatile data to be saved, which consumes more power during the store and restore process, and occupies more non-volatile memory. Second, the size of the decoupling capacitor is limited and fixed, so the stored energy may not be enough to complete a checkpointing process.

\subsubsection{Task-based Checkpointing}
To support intermittent program execution, there exist works that decompose a long-running program into a sequence of short and atomic tasks, such as sampling a sensor reading. Furthermore, the idea of \textit{idempotency}~\cite{de2012static,de2013idempotent} was employed to address the data inconsistency issue. Specifically, if a task is idempotent, it can be executed multiple times without producing different results. To ensure idempotency, the task should not contain any \textit{idempotency violation}, which is a write instruction to a non-volatile memory that was first accessed by a read instruction, termed as \textit{write-after-read} (WAR). Recall the example in Figure~\ref{fig:data-consistency}, non-volatile variable \textit{Encrypted\_Glucose} is first read after the checkpoint and there is another line of code \textit{Encrypted\_Glucose = 1*2} assigning a new value to it before the next checkpoint, i.e., a WAR. As long as a power failure happens after the write instruction, a power failure before the next checkpoint would result in data inconsistency.

Depending on how these tasks are connected, these works can be further divided into two categories. First, the decomposed tasks are connected by lightweight checkpoints, such as DINO~\cite{lucia2015simpler}, Clank~\cite{hicks2017clank} and Ratchet~\cite{van2016intermittent}, in which a checkpointing process would be launched at the task boundaries. DINO and Ratchet analyzed the idempotency and decomposed a program at compile-time using static code analysis~\cite{hofmann2003static}, which imposes heavy burdens on the programmers.  The tool CleanCut~\cite{colin2018termination} can automatically decompose programs into efficient, idempotent tasks at compile-time. At run time, the executions were performed task-by-task and a checkpointing process is launched after a task. In contrast, Clank dynamically and automatically decomposed program executions into a stream of sections at run-time. 

\begin{figure}[t]
	\centering
	\centering
		\includegraphics[scale=0.52]{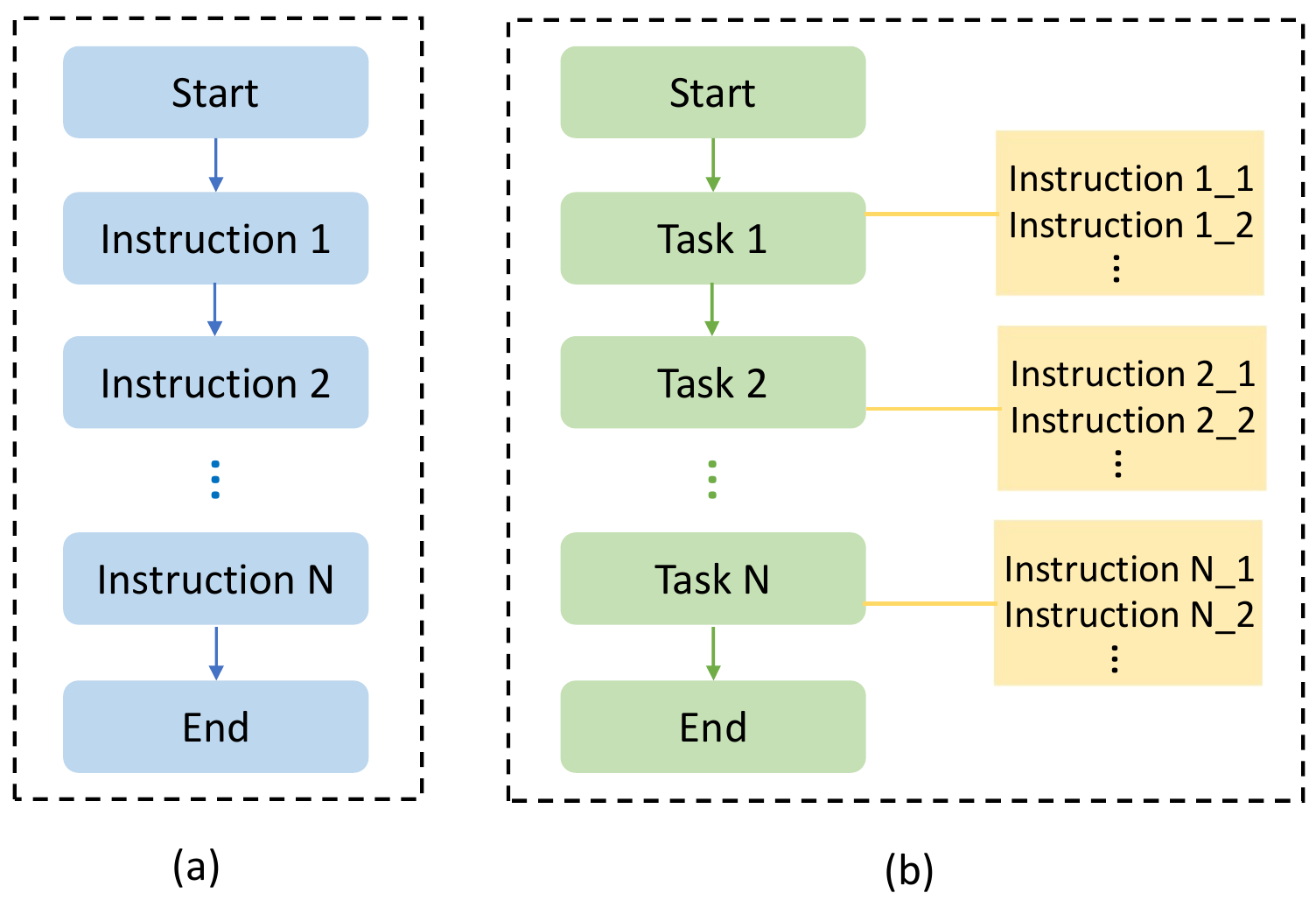}
	\caption{Comparison of (a) conventional line-based coding  and (b) task-based programming model.}
	\label{fig:task-model}
\end{figure}

\begin{figure}[t]
	\centering
	\centering
		\includegraphics[scale=0.46]{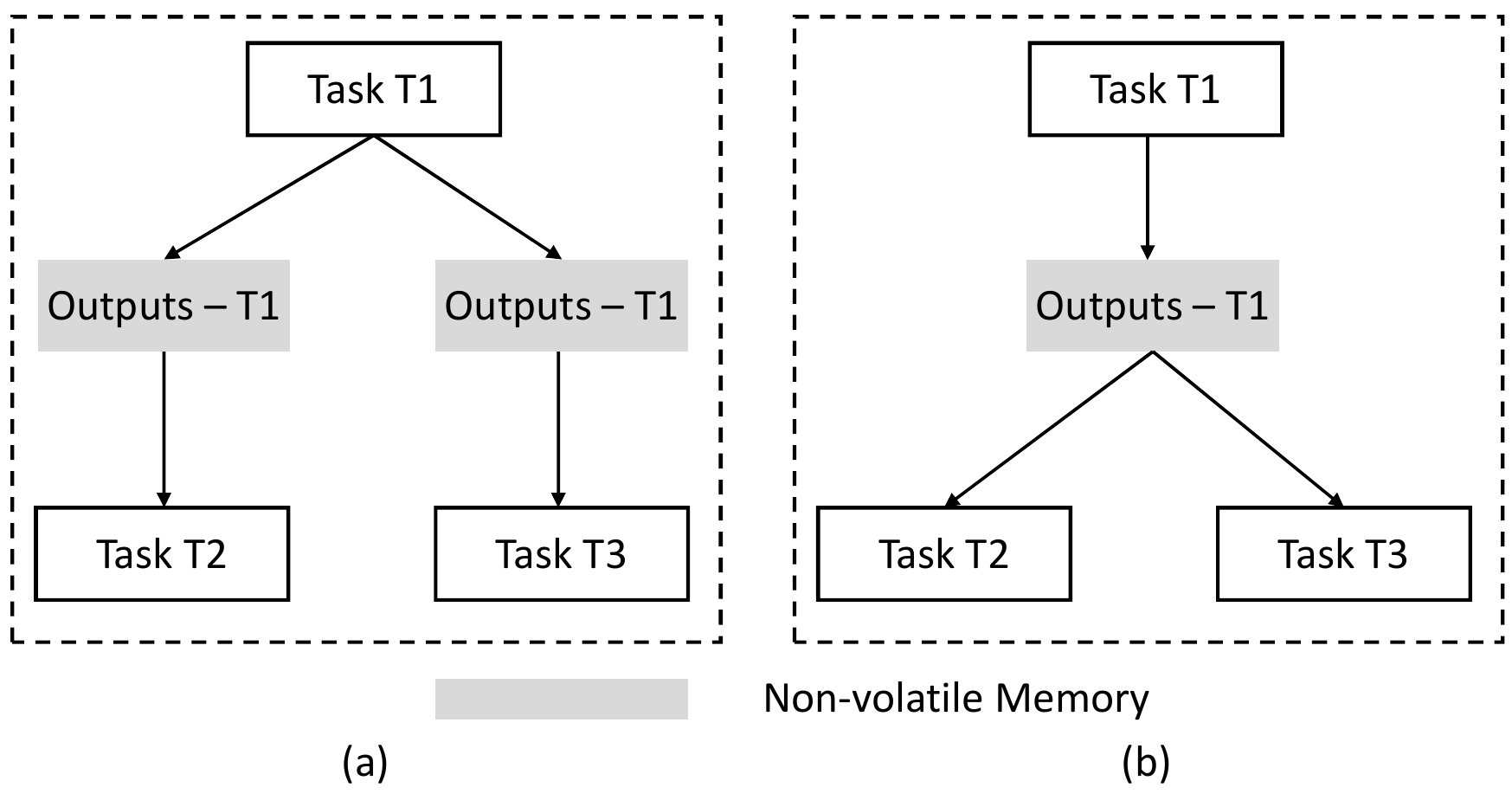}
	\caption{Comparison of data exchange mechanism in (a) Chain and (b) Alpaca.}
	\label{fig:data-exchanging}
\end{figure}

Second, the execution just follows a task flow and no checkpointing is needed between tasks, such as Chain~\cite{colin2016chain} and Alpaca~\cite{maeng2017alpaca}. Instead of decomposing a long program that is written line-by-line (or instruction-by-instruction), Chain and Alpaca designed a new programming model where a program is written at the granularity of a task, i.e., groups of instructions, and these tasks are connected through a control flow graph. Figure~\ref{fig:task-model} illustrates the comparison of conventional line-based coding model and task-based programming model. In addition, Chain and Alpaca proposed a judicious data management scheme that eliminates checkpointing as well as data inconsistency. In detail, the variables are defined as either task-local or task-shared, where the former is limited to use in individual tasks and stored in volatile memory, while the latter is defined in the global scope and stored in non-volatile memory. Since a task is idempotent and its inputs are stored in non-volatile memory, it is expected to produce the same results from power failures, thereby preserving data consistency. 
For task-shared variables, Chain allocated a block of non-volatile memory (termed as \textit{channel}) to \textit{each pair} of tasks, which is not memory efficient as it creates multiple versions of a task's inputs/outputs. Alpaca solved this problem by discarding \textit{channels} and linking the memory blocks of different tasks directly. Figure~\ref{fig:data-exchanging} presents the comparison of these two data exchange mechanisms.

\subsubsection{Watchdog Checkpointing}
In the checkpointing approach, if the maximum energy budget of the storage capacitor is not enough to support the executions to reach the next checkpoint, the instructions from the last checkpoint would be executed repeatedly and never completed. This is also applicable to the task-based method when the maximum available energy is not enough to successfully accomplish a task, e.g., when a task is too long. To deal with the `Sysiphean' problem and guarantee program progress, a watchdog timer has been widely adopted in the literature~\cite{van2016intermittent,ransford2012mementos,mirhoseini2013idetic,bhatti2017harvos,hicks2017clank}. Specifically, when the program re-executes from a checkpoint or a task boundary, a watchdog timer is initiated. Once the timer interrupt occurs, it launches a checkpointing process to mandatorily split previously uncompleted code group into two or multiple smaller groups. Although watchdog checkpointing is not optimized in terms of the energy and memory overhead, it is the most effective way to address the `Sysiphean' problem.

\subsection{Timekeeping across Power Failures}
Maintaining a reliable sense of time in battery-powered devices can be easily achieved using an internal clock (e.g., RTC or real-time clock). For the intermittently-powered devices, however, it is very challenging as the system clock is turned off when the power fails. Thus, to execute a meaningful task, the first challenge is to track the time elapsed between power failures. In~\cite{rahmati2012tardis,hester2016persistent}, the authors exploit \textit{remanence decay} of SRAM (Static Random-Access Memory, a type of volatile memory) and capacitor for timekeeping, respectively. 

The principle of remanence decay on SRAM is that the SRAM cells lose their states (reset from 1 to 0) \textit{gradually} when the power is cut off. The next time the device is powered on, the elapsed time can be roughly estimated based on the percentage of cells remaining 1. Similarly, the idea of remanence decay on capacitor is based on the fact that a capacitor voltage dissipates slowly after being disconnected from a power supply. Thus, the elapsed time can be estimated by measuring the capacitor voltage after reboot. TARDIS~\cite{rahmati2012tardis} is a timekeeper based on SRAM decay, which provides coarse-grained time tracking. CusTARD~\cite{hester2016persistent} is based on the capacitor voltage decay and allows finer-grained timekeeping, at the cost of equipping an additional capacitor on the device. Either of the timekeepers can accurately track time through power failures up to 45 secs. A detailed comparison of TARDIS and CusTARD can be found in~\cite{hester2016persistent}.

After guaranteeing a sense of time, i.e., the time elapsed between power failures is obtained, it is critical to efficiently determine whether to execute from previous states or to launch a new task. In~\cite{hester2017timely}, the authors presented Mayfly, a programming language as well as a run-time system that supports timely execution on intermittently-powered devices. As a programming language, Mayfly programs consist of a sequence of tasks, like Chain and Alpaca. The difference is that, to take the timing information into consideration, it designs additional attributes to each task, such as priority, expire time, and the \textit{collect} (a command that assigns the amount of data required). Mayfly exploits the CusTARD timekeeper to track the time elapsed through power failures. After reboot, the Mayfly runtime first estimates the elapsed time from last power outage and updates the local system time. Then, according to the current system time, the lifetime of previous data can be calculated. If the data are expired, it discards the data and rollback to the beginning of the task. Otherwise, the execution continues. 

\subsection{Lessons Learned}

From recent studies of EH powered IoT systems, we learn the following key lessons:
\begin{itemize}
\item Without a stable energy supply, the intermittent, unpredictable, and uncontrollable energy arrival poses unique challenges for on-board program execution in EH-IoTs. To guarantee program correctness and efficient execution, novel program execution models that can adapt to intermittent computing are required.

\item Conventional checkpointing mechanism is unable or inefficient to deal with power failures for batteryless EH-IoTs. Various techniques have veeb developed, such as advanced ADC interrupt and idempotency analysis, to optimize the checkpointing process. Most recently, specialized programming languages and execution models for intermittent computing have been devised and shown excellent energy utilization efficiency.

\item Timekeeping, i.e., maintaining the system clock during power failures, is critical to ensure timing consistency of computation (e.g., freshness of sensor data). However, current timekeeping techniques can only track the elapsed time within one minute. In case of energy-starving scenarios that may suffer from long time power failures, novel timekeeping techniques are required.
\end{itemize}

\section{Communications for EH-IoTs}
\label{section:communication}

There exist different approaches in the literature to design efficient communication solutions that can cope with tiny and unpredictable energy harvested in small form factor IoT sensors. Although several articles~\cite{ku2016advances,ahmed2015survey,ulukus2015energy} have surveyed this well-investigated topic, some emerging approaches have not been examined adequately. For the benefit of the readers, in this section we first provide a comprehensive taxonomy  of energy harvesting communications research, and identify topics that require further investigation (Section \ref{s:tax}). Then we survey those less-explored topics. In particular, we survey three communications approaches relevant for EH-IoTs: on-line optimization of transmission powers using reinforcement learning (Section \ref{s:RL}); packet-less communication protocols (Section \ref{s:packetless}); and advanced reflective (backscattering) communication techniques (Section \ref{s:IRS}).

\subsection{Taxonomy of Energy Harvesting Communications Research}
\label{s:tax}

\begin{figure}[t]
	\centering
	\centering
		\includegraphics[scale=0.55]{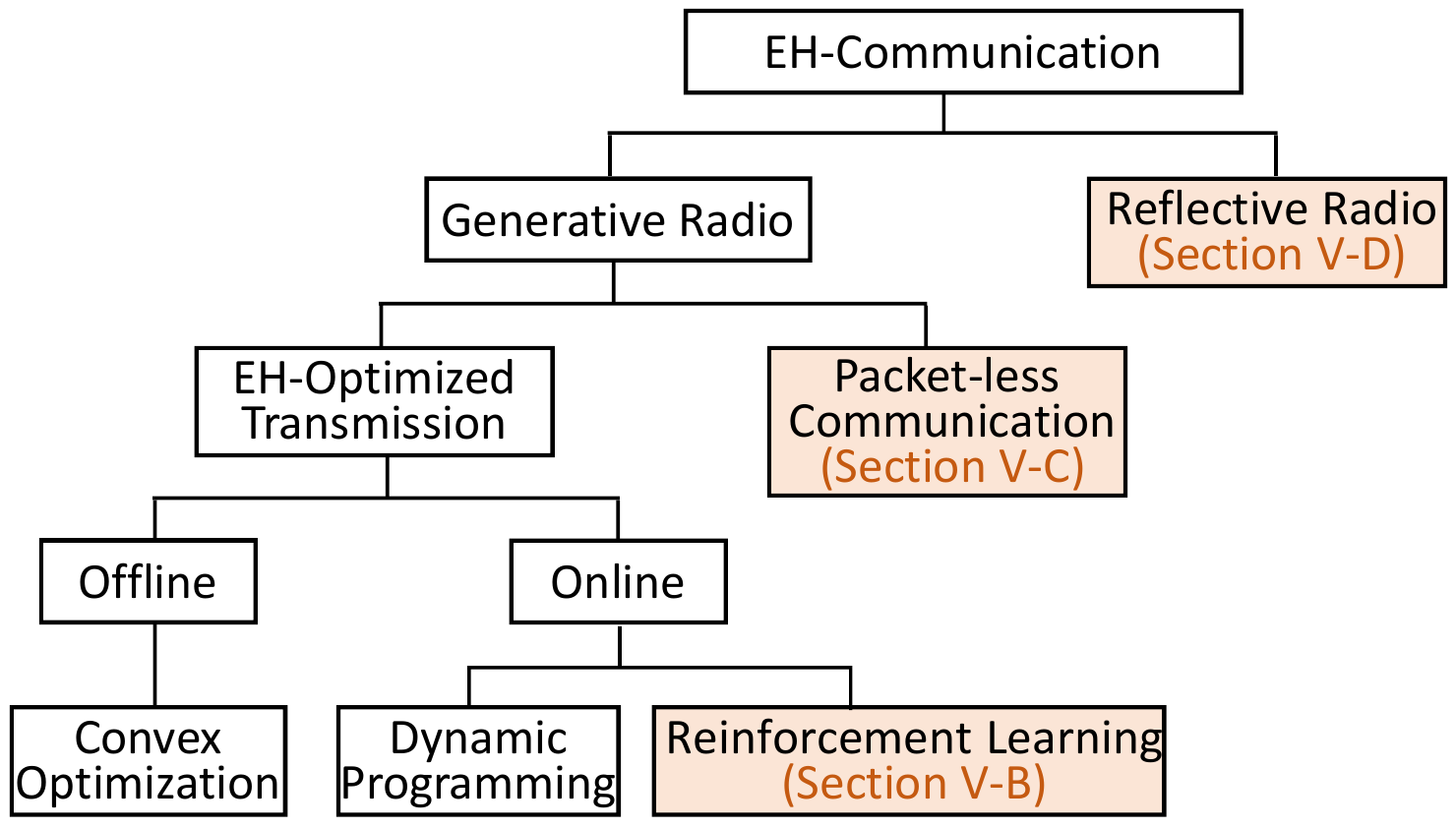}
	\caption{Taxonomy of energy harvesting communications.}
	\label{fig:EH_com_taxonomy}
\end{figure}

Figure~\ref{fig:EH_com_taxonomy} provides a taxonomy of energy harvesting communications research conducted in recent years. Fundamentally, all IoT wireless transmissions can be categorized into two broad types based on how the signal waves are transmitted by the device. The first type is \textit{generative radio}, which requires the transmitter to actively generate RF waves to carry information. In contrast to generative radio, the \textit{reflective radio} transmits data by modulating and reflecting RF waves that impinge on its surface.

Research in generative-radio-based communications have two distinct flavors. The mainstream research, which we refer to as \textbf{EH-optimized Transmission}, considers conventional packet-based transmissions, but strives to optimize the transmission schedule and power of the packets as a function of the current channel and energy harvesting states. The optimization problem has been extensively studied and surveyed~\cite{ku2016advances,ahmed2015survey,ulukus2015energy}. Table~\ref{tab:communication summury} summarizes these optimization studies under different networking contexts, which reveals that, irrespective of the networking contexts, the researchers basically considered two different optimization approaches -- offline or online. The offline method assumes that the transmitter has perfect \textit{a priori} knowledge of the communication channel as well as the energy arrival process, which can be solved by convex optimization techniques \cite{ku2016advances}. 

Offline optimization is less practical but provides insight that can be used to design a good practical solution. Online optimization, on the other hand, does not have to know the channel or energy arrivals and can make the power allocation decisions based only on the currently observed channel and energy states. The optimization can be solved using dynamic programming if the statistical knowledge of the underlying system model is known. However, for many practical IoT applications, it is difficult to gather statistics on energy harvesting prior to deployments. In such cases, \textit{reinforcement learning} can be very effective to learn the system model online and derive the optimal power allocation policy without having to know the transition probabilities. Previous work published until 2016 \cite{ku2016advances,ahmed2015survey,ulukus2015energy,lu2015wireless,he2015survey} surveyed only a few reinforcement learning based power allocation optimizations. Since then, reinforcement learning has gained significant attention and many more papers have been published since 2016. Therefore, in Section~\ref{s:RL}, we provide a brief survey of recently published works on reinforcement learning applied to communication optimization in EH-IoTs.

\textbf{Packet-less transmission} is another interesting design paradigm to minimize energy consumption for generative radios. The basic idea is to convey the detection of certain events by transmitting a single pulse (or bit) without using elaborate packet headers. Works on packet-less communications are reviewed in Section \ref{s:packetless}.

Finally, there is a growing body of research related to reflective radio, also known as backscattering. As backscattering does not have to generate an original signal from scratch, it promises to dramatically reduce wireless transmission energy cost of EH-IoTs. Communication with reflective radio, however, poses many challenges as it heavily depends on the availability of impinging radio waves when it needs to communicate. How to efficiently modulate an impinging wave to achieve high data rate is another major focus of research. Although such research started several decades ago with the basic radio frequency identification (RFID) systems, it has evolved significantly over the years. The most recent trend is to use some type of meta-surface or Intelligent Reflective Surface (IRS) containing high density of software controlled tiny reflective units, which can be programmed to dynamically control the amplitude, phase, and frequency of the reflected wave. Although previous surveys have covered the basic RFID backscattering~\cite{khan2009survey} as well as the more ambient form of backscattering~\cite{van2018ambient}, they did not survey the works that involve IRS. Section \ref{s:IRS} surveys the IRS-based radio reflection research.

\begin{table*}[t]
\centering
\caption{Examples of EH-communication optimization problems.}
\label{tab:communication summury}
\setlength\tabcolsep{9.8pt}
\ra{1.2}
\begin{tabular}{@{}lll@{}}\toprule

\textbf{Network Context} &\textbf{Problem/Objective}  &\textbf{Strategy}   \\ \midrule

\multirow{4}{*}{Point-to-point} &Optimal power allocation to maximize throughput~\cite{he2014recursive}    & Offline, Convex optimization \\ \cmidrule{2-3}

& Optimal energy management  to minimize energy consumption~\cite{prabuchandran2013q}    & Online, MDP, Reinforcement learning \\ \cmidrule{2-3}

& Optimal transmission policy to maximize throughput~\cite{blasco2013learning}    & Online, MDP, Reinforcement learning \\ \cmidrule{1-3}
   
\multirow{4}{*}{Cooperative Network} &Optimal power and rate allocation at Tx/relay to maximize throughput~\cite{huang2013throughput}   & Offline, Convex optimization \\ \cmidrule{2-3}
 
&Optimal relay selection to minimize transmission power~\cite{luo2016transmit}   & Offline, Convex optimization  \\ \cmidrule{2-3}

&Optimal relay scheduling to maximize throughput~\cite{li2011relay}   & Online, PO MDP, Dynamic programming  \\ \cmidrule{1-3}

\multirow{3}{*}{Cognitive Network} & Optimal SU channel access policy to maximize throughput~\cite{yin2015achievable} &   Offline, Convex optimization   \\ \cmidrule{2-3}
 
&Optimal power allocation for cognitive relay to maximize throughput~\cite{yin2014optimal}   & Offline, Convex optimization  \\ \cmidrule{2-3}

&Optimal SU channel access policy to maximize throughput~\cite{hoang2014opportunistic}   & Online, MDP, Reinforcement learning  \\ \cmidrule{1-3}

\multirow{3}{*}{Cellular Network} &Optimal user scheduling to maximize throughput~\cite{cui2012delay}    & Online, PO MDP, Dynamic programming   \\ \cmidrule{2-3}
 
& Optimal BS ON/OFF decision to maximize its availability region~\cite{dhillon2014fundamentals}   & Offline, Convex optimization     \\ \cmidrule{1-3}

\multirow{3}{*}{WIPT} &Optimal transmission beamforming to minimize power consumption~\cite{shi2014joint}   & Offline, Convex optimization   \\ \cmidrule{2-3}

&Optimal power splitting scheme to maximize power transfer and throughput~\cite{liu2013wireless}   & Offline, Convex optimization   \\ 
 \bottomrule 
 \multicolumn{3}{l}{ *MDP - Markov Decision Process, PO MDP - Partially Observed Markov Decision Process, WIPT - Wireless Information \& Power Transfer. }
\end{tabular}
\end{table*}

\begin{table*}[t]
\centering
\caption{Summary of recent advancements using reinforcement learning in EH-IoTs.}
\label{tab:RL summury}
\ra{1.2}
\setlength\tabcolsep{7.7pt}
\begin{tabular}{lllll}
\toprule
   \textbf{Network}&\textbf{\multirow{2}{*}{Issue}} &\multirow{2}{*}{\textbf{Reference}}  &\textbf{Perform } &\textbf{Methodology} \\  
   
   \textbf{Context}&&  &\textbf{RL at} & \\ \midrule
 &\multirow{2}{*}{Power allocation}& \cite{ortiz2016reinforcement}   & Tx&  Continuous MDP, linear approximation, binary functions, and SARSA \\ \cmidrule{3-5}

 &  & \cite{sakulkar2018online,kim2018action}   & Tx&  Discrete MDP, LPSM, action-bounding deep Q-learning \\ \cmidrule{2-5}

Point-to-point & Transmission decision & \cite{wu2017optimal}   & Tx&  Continuous MDP, polynomial approximation, after-state SARSA  \\ \cmidrule{2-5}

 & Transmission policy & \cite{ayatollahi2017reinforcement}   & Tx&  Discrete MDP, Q-learning  \\ \cmidrule{2-5}

 & Exploration \& exploitation & \cite{masadehreinforcement}   & Tx&  Discrete MDP, SARSA  \\ \cmidrule{1-5}
 
\multirow{5}{*}{WSN}& Energy prediction & \cite{kosunalp2016new} & Nodes&  Q-learning with real solar data  \\ \cmidrule{2-5}
 
& Sleep scheduling & \cite{chen2016reinforcement}   & Nodes&  Q-learning with real solar data  \\ \cmidrule{2-5}
  
&\multirow{2}{*}{Power management} & \cite{shresthamali2017adaptive,hsu2018fuzzy}   & Nodes&  Q-learning\cite{shresthamali2017adaptive}, Fuzzy Q-learning\cite{hsu2018fuzzy} with real solar data  \\ \cmidrule{3-5}

&  & \cite{aoudia2017learning}   & Nodes& Discrete MDP, actor-critic learning algorithm  \\ \cmidrule{1-5}

\multirow{4}{*}{Cellular}& Access control& \cite{chu2018reinforcement,chu2018reinforcementnew}  & BS & Discrete MDP, LSTM deep Q-learning  \\ \cmidrule{2-5}
 
& BS ON/OFF switching & \cite{miozzo2017switch}   & BS& Discrete MDP, distributed multi-agent Q-learning  \\ \cmidrule{2-5}
  
&Resource allocation & \cite{wei2018user}   & BS & Continuous MDP, linear approximation, actor-critic learning algorithm \\ \cmidrule{1-5}

\multirow{3}{*}{WIPT}& \multirow{3}{*}{Transmission interval/rate}& \cite{li2018reinforcement}  & Rx & Discrete MDP, Q-learning  \\ \cmidrule{3-5}
 
&  & \cite{chun2018adaptive}   & Tx/Rx& Discrete MDP, Q-learning  \\ \cmidrule{1-5}

Cooperative&  Power allocation&\cite{ortiz2017reinforcement}   & Tx/Relay & Continuous MDP, linear approximation, SARSA  \\ \midrule

Cognitive & SU action decision  &\cite{wu2019sensing}   & SU & Discrete MDP, after-state, policy-based learning algorithm  \\ \midrule

\multirow{3}{*}{MEC}& \multirow{3}{*}{Data offloading}& \cite{xu2017online}  & Edge & Discrete MDP, post-decision state based learning  \\ \cmidrule{3-5}
 
&  & \cite{min2018learning,min2019learning}   & IoTs& Discrete MDP, post-decision state based learning, deep CNN  \\

\bottomrule 
\multicolumn{5}{l}{SU - Secondary User, MEC - Mobile Edge Computing.}
\end{tabular}
\end{table*}

\subsection{Reinforcement Learning based Communication Optimization in EH-IoTs}
\label{s:RL}
Of late, application of reinforcement learning to communication optimization in EH-IoTs is receiving significant interest as it requires minimal system state information and is suitable to implement in practical scenarios. A variety of issues, e.g., transmission power allocation, transmission policy, and user scheduling, under different network contexts have been investigated. Based on the number of considered system states, the optimization problem can generally be modeled as continuous Markov Decision Process (MDP) and discrete MDP depending on if the system state is infinite and finite, respectively. When solving the problems, the continuous MDP requires one more step to handle the infinite states, and techniques like linear approximation~\cite{ortiz2016reinforcement,wei2018user,ortiz2017reinforcement} and polynomial approximation~\cite{wu2017optimal} are proposed. In the following, we survey the reinforcement learning approaches proposed since 2016 under different networking contexts with a succinct summary presented in Table~\ref{tab:RL summury}. 

{\bf Point-to-point communications:} In the point-to-point EH communications, reinforcement learning has been extensively explored. Transmission power allocation was investigated in~\cite{ortiz2016reinforcement,sakulkar2018online,kim2018action}. Specifically, in~\cite{ortiz2016reinforcement}, continuous MDP was formulated by exploiting linear approximation and binary functions to handle infinite states and state-action-reward-state-action (SARSA) algorithm to learn the optimal policy. In contrast, in~\cite{sakulkar2018online} discrete MDP was used to propose the linear program of sample means (LPSM) algorithm to learn the optimal power allocation policy. The authors in~\cite{kim2018action} proposed an action bounding deep Q-learning algorithm to accelerate the learning process; while in~\cite{wu2017optimal} the transmission decision problem was investigated to determine whether to transmit a packet or not, followed by an after-state SARSA learning algorithm. In~\cite{ayatollahi2017reinforcement}, the authors considered a multiple input and multiple output (MIMO) system where the transmitter can change the number of antennas during transmission, and employed Q-learning to learn the optimal transmission policy. The SARSA algorithm was utilized in  ~\cite{masadehreinforcement} to investigate the exploration and exploitation balancing problem; it is demonstrated that the convergence-based algorithm outperforms the epsilon-greedy algorithm. 

{\bf Wireless sensor network (WSN):} Some literature investigated the implementation of reinforcement learning in EH WSN, using real-world experiment data. Wireless sensor nodes equipped with a solar cell perform sensing tasks and transmit the information to a remote server. Kosunalp et al.~\cite{kosunalp2016new} utilized Q-learning to predict the solar energy arrival and the experimental results demonstrated that a prediction error ratio of 0.3 was achieved. Chen et al.~\cite{chen2016reinforcement} considered the optimal policy that schedules the state (active or sleep) of sensor nodes to maintain effective area coverage. Q-learning algorithm was implemented and the experiment results suggested that a coverage ratio of 0.99 can be obtained. Paper~\cite{hsu2018fuzzy},\cite{shresthamali2017adaptive} and~\cite{aoudia2017learning} considered the power management policy at the sensor nodes, i.e., how to schedule the energy for sensing, transmission and sleeping. Hsu et al.~\cite{hsu2018fuzzy} proposed and implemented a fuzzy Q-learning algorithm, while Shresthamali et al.~\cite{shresthamali2017adaptive} implemented the Q-learning algorithm. Without real-world experiments, Aoudia et al.~\cite{aoudia2017learning} modeled the problem as a discrete MDP and proposed an actor-critic learning algorithm.

{\bf Cellular networks with small cells:} To increase capacity and spectral reuse, future cellular networks are expected to deploy small cells controlled by small form factor base stations (BSs).  It is possible to power these small BSs as well as the user equipments that connect to them with energy harvested from the environment. However, due to the stochastic nature of the energy source, incorporation of energy harvesting into both BS and user equipment poses more challenges on managing the network, such as user access control~\cite{chu2018reinforcement,chu2018reinforcementnew}, BS ON/OFF switching~\cite{miozzo2017switch}, and resource allocation~\cite{wei2018user}. In~\cite{chu2018reinforcement} a long short-term memory (LSTM) deep Q-learning algorithm is proposed to learn the optimal multi-access control policy at the BS. Using the proposed method, in a followup work~\cite{chu2018reinforcementnew}, the authors investigated joint access control and battery prediction problem, and developed a two-layer reinforcement learning network to maximize the sum rate and minimize the prediction loss. Assuming the traffic of small cell BS can be offloaded to a macro BS, in~\cite{miozzo2017switch} the optimal ON/OFF switching policy of small cell BSs was investigated to improve energy efficiency of the overall network. Specifically, distributed multi-agent Q-learning was proposed to learn the energy income and traffic demand patterns at each small cell BS and determine its ON/OFF state. Similarly, to improve network energy efficiency, the authors in~\cite{wei2018user} utilized the actor-critic algorithm to investigate the optimal policy for user scheduling and resource allocation.

{\bf Wireless information and power transfer (WIPT):} WIPT allows the transmitter and receiver to exchange energy and information in a time-devision manner. To achieve the optimal system performance, time slot allocation for information and energy transmission is critical. Recent works~\cite{li2018reinforcement,chun2018adaptive} attempted to solve this problem using reinforcement learning. Specifically, the authors in \cite{li2018reinforcement} considered that the Tx transmits energy to the Rx, and the Rx uses the harvested energy to send data back to the Tx. The Q-learning algorithm was exploited to find the optimal data transmission rate at the receiver. The results domonstrated that the algorithm can reduce packet loos rate by 60\%, compared to the random selection scheme. In contrast, in~\cite{chun2018adaptive}, it was considered that the Tx transmits both information and power to the Rx. To maximize the receiption rate at the Rx, the Q-learning algorithm was utilized to learn the optimal data transmission rate at the Tx and energy harvesting interval at the Rx.  

\begin{table*}[t]
\centering
\caption{Summary of methods for packet-less event notification using a single pulse in EH-IoTs.}
\label{tab:iont summary}
\ra{1.2}
\begin{tabular}{m{1.5cm}<{}m{2.7cm}<{}m{2.5cm}<{}m{1.7cm}<{}m{5.8cm}<{}m{1.2cm}<{\raggedright}}\toprule
   \textbf{Ref}&\textbf{Application} &\textbf{Pulse medium} &\textbf{Detectable \# of event types} & \textbf{Method of detection}  &\textbf{Energy source} \\  \midrule
\cite{das2017towards} & SHM for aeroplane& ultrasound through metal substrate & binary  &  DoA for location, pulse for event & wing vibrations   \\ \midrule

\cite{zarepour2017semonnew} & IoNT for chemical reaction detection& THz via air& $>$ binary  &  pulse amplitude for classification at AP, pulse width for location, pulse energy for event & events   \\ \midrule

\cite{hassan2019eneutral} & IoNT event monitoring& THz via air& $>$ binary  &  pulse energy for event (location cannot be detected) & chemical reactions   \\ \midrule

\cite{shree2018classifying,prasad2018direction} & IoNT event monitoring& THz via air& $>$ binary  &  DoA for location, derivative order for event & anything   \\ \midrule

\cite{prasad2018energy} & IoNT event monitoring& mmWave/THz via air& $>$ binary  &  DoA for location, central frequency for event & anything   \\ 

 \bottomrule 
 \multicolumn{6}{l}{ *SHM - Structure Health Monitoring.}
\end{tabular}
\end{table*}

{\bf Cooperative and cognitive networks:} In~\cite{ortiz2017reinforcement}, reinforcement learning is employed to find the optimal power allocation policy for Tx and relay in cooperative EH networks, with the purpose of maximizing the throughput at the receiver. Specifically, two-hop communication was split into two point-to-point problems (the transmitter to the relay and the relay to the receiver) and solved by the method individually ~\cite{ortiz2016reinforcement}. In cognitive networks, the secondary user (SU) can perform either spectrum sensing (to detect whether the spectrum is free), channel probing (to acquire channel state information, or data transmission (need to configure the transmission power) at each time slot. Assuming the SU is powered by the energy harvested from the ambient environment, in ~\cite{wu2019sensing} a policy-based learning algorithm is proposed to determine the SU's action with the goal of maximizing its throughput.  

{\bf Mobile edge computing (MEC):} In (MEC) systems, the end IoT devices offload data to the edge servers, which offload part of its workload or preprocessed data to a remote cloud. Considering EH MEC where both the IoT devices and edge servers can harvest energy from the environment, reinforcement learning has been applied to optimize the offloading policy at IoT devices and edge servers. At the edge servers, a post-decision state (PDS) based learning algorithm is proposed in~\cite{xu2017online} to obtain the optimal workload offloading (to the centralized cloud) policy, which minimizes the long-term system cost. Instead, at the IoT devices, the PDS based learning and deep convolutional neural network (CNN) are combined in~\cite{min2018learning,min2019learning} to select the optimal edge device and offloading rate. 

\subsection{Packet-less Communication}
\label{s:packetless}
In many IoT applications, the deployed sensor nodes only need to detect an event, such as normal vs. faulty operation of the monitored structure, and transmit that information to a nearby access point. Because the event information is either just binary or event identification from a small set of possible event types, a single pulse could be used to convey this information, thus eliminating the need to transmit a packet with its associated overhead, such as node identification address, preamble, synchronization, etc. In low energy harvesting environments, this can help maintain the energy consumption of the device below the energy production rate.

Such packet-less pulse-based event notification has been investigated by different research groups. Without packet header and payload, the main issues to solve is how to localize the event and how to convey the event type from a single pulse. For structural health monitoring of an airplane wing, the authors in~\cite{das2017towards} proposed transmitting an ultrasonic pulse from a sensor to a nearby access point through the metal substrate to convey the detection of fault when vibration exceeds a given threshold. The sensor nodes are placed in a carefully designed cellular diagram, which allows the access point to detect the position of the event by detecting the direction of arrival of the pulse. The vibrations are also used to harvest energy and power the pulse, thus eliminating the need for batteries. 

The idea of using a single pulse to convey event information is particularly attractive for nanoscale IoT, also known as Internet of Nano Things (IoNT), where nanoscale sensors must conserve transmission energy as much as possible to maintain energy consumption below the extremely limited amount of energy it can possibly harvest from the environment. A detailed analysis in \cite{canovas2018nature} revealed that even with 1 pJ energy consumption per pulse, the transmission power would account for 68\% of the total device power consumption, suggesting that the number of pulses must be minimized to achieve a balance between power consumption and generation in IoNT. In a recent work \cite{hassan2019eneutral}, we have shown that if the events to be detected emit distinct amounts of energies, it is possible to convey both event type and location information using a single pulse. This is possible if all harvested energy from the event is used to transmit the pulse with a carefully allocated pulse width for each sensor node located at different locations. In this case, the amplitude of the pulse is influenced by both the type of event and its location, thus a classifier at the access point can uniquely detect the event type and its location by simply detecting the amplitude of the pulse. 

The method proposed in \cite{hassan2019eneutral} does not scale to large number of sensor nodes as the number of classes to classify using only the amplitude of the pulse increases. In~\cite{shree2018classifying,prasad2018direction,prasad2018energy}, we have further shown that if the access point can be configured with necessary hardware for detecting the direction of arrival (DoA) of the pulse, then a given pulse feature only has to classify the event type as the location is already obtained by DoA. We have demonstrated that as higher derivative of Gaussian pulses are often used for IoNT, we can use a range of pulse features for detecting event types, including the derivative order \cite{shree2018classifying,prasad2018direction} as well as the center frequency \cite{prasad2018energy} of the Gaussian pulse. Table \ref{tab:iont summary} compares various methods of conveying event notification using a single pulse.

\subsection{Reflective Radio Communications}
\label{s:IRS}
This section first discusses the evolution in reflective radio technology and then surveys some of the very recent works involving IRS.

\subsubsection{Reflective Radio Evolution}
Power consumption of an IoT sensor is typically dominated by wireless communications because conventional radio modules (e.g., Wi-Fi and Bluetooth) are designed to actively generate new radio waves (RF signals) each time some data needs to be transmitted. Instead of generating RF signals, a reflective radio transmitter sends data by simply modulating and reflecting the incident RF signals. By eliminating the power consuming RF generation elements from the radio circuit, the reflective radio enables small IoT sensors to operate even with a tiny amount of harvested power. The principle of reflective radio or backscattering was first commercialized in 1983 in the form of batteryless RFID, which saw explosive adoption in the subsequent years. Millions of RFID tags are currently in operation to track goods, machines, animals, and many other types of objects.

Although RFID is very popular, it has several drawbacks. RFID tags require a carrier emitter (a reader) to transmit electromagnetic waves from a short distance so that the tags can reflect the waves back to the reader. The requirement of readers limits the pervasive deployments of RFID. Another major limitation of the RFID technology is that the tags can communicate only with the reader using the same radio frequency used by the reader. To overcome these limitations, in around 2013, the basic backscattering evolved into the so called \textit{ambient backscattering}~\cite{wang2016ambient,van2018ambient}, which is capable of modulating and reflecting the surrounding RF signals that are transmitted by ambient RF sources, such as TV towers, cellular base stations, and Wi-Fi access points.  Ambient backscattering thus enables direct communication between two nodes allowing a distributed communication network. Another major advantage of ambient backscattering is that it can piggyback the IoT data on the licensed spectrum used by TV, cellular etc., thus alleviating pressure on the spectrum.

\begin{figure}[t]
	\centering
	\centering
		\includegraphics[scale=0.8]{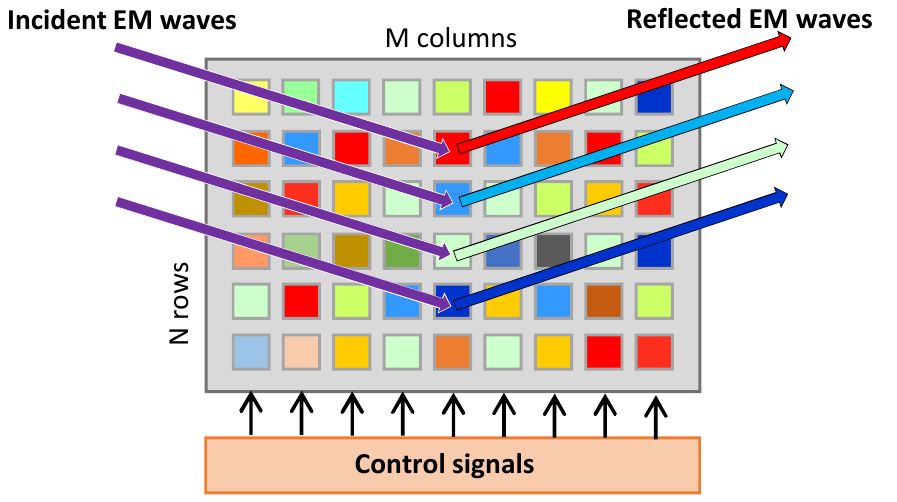}
	\caption{Schematic of intelligent reflective radio surface (IRS).}
	\label{fig:irs}
\end{figure}

\begin{table}[t]
 \centering
 \caption{Comparison of three reflective radio techniques: RFID, Ambient Backscattering, and Intelligent Reflective Surface (IRS).}
 \label{t:evolution}
 \setlength\tabcolsep{1.4pt}
 \ra{1.2}
 \begin{tabular}{ccccc}
 \hline
 
\textbf{Reflective} &\textbf{Use Ambient }& \textbf{Programmability }&\textbf{Communication} & \textbf{Milestone} \\ 
 \textbf{Radio} &\textbf{EM Waves} & \textbf{of Reflection} & \textbf{Architecture}& \textbf{Year} \\ \hline
  
RFID & No  & No &Centralized& 1983   \\ \hline
 
  Ambient & \multirow{2}{*}{Yes}  & \multirow{2}{*}{No} &\multirow{2}{*}{Distributed} & \multirow{2}{*}{2013}   \\ 
 
 Backscatter &  &  & &    \\ \hline
 
IRS & Yes  & Yes &Distributed & 2018  \\ \hline   
 \hline

 \end{tabular}
 \end{table}

\begin{figure}[t]
	\centering
	\centering
		\includegraphics[scale=0.59]{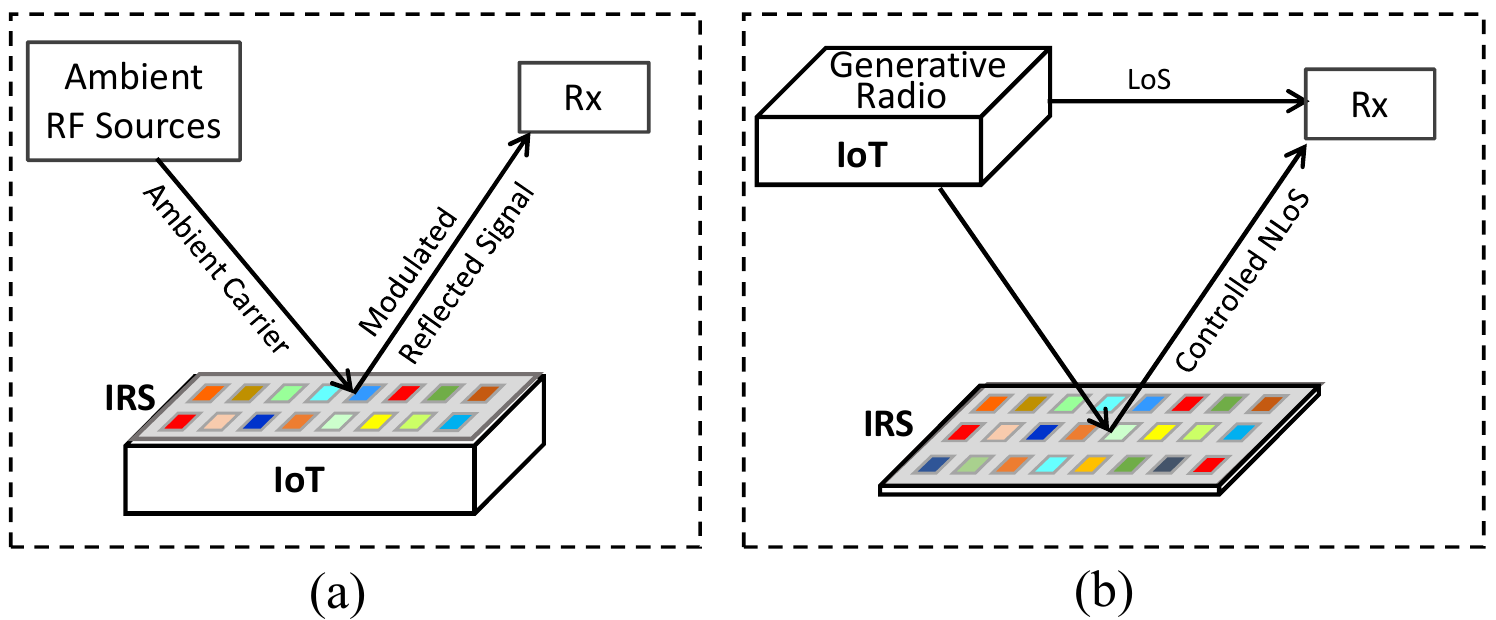}
	\caption{IRS use cases for IoT: (a) Enhanced ambient backscattering with programmable reflection and advanced modulation of reflected signal (IRS deployed within the IoT), and (b) IRS-assisted IoT communication with controlled multipath propagation (IRS deployed in the environment).}
	\label{fig:irs_arc}
\end{figure}

Most recently, a novel reflective radio technology, referred to as intelligent reflective surface (IRS), has gained massive attention~\cite{liang2019large,di2019smart}. Unlike RFID and ambient backscattering that have a pre-determined reflection property, the IRS is made from an artificial structure of 2D elements (see Figure~\ref{fig:irs}) whose reflection properties, such as refraction, absorption, and reflection can be adjusted independently and electronically \textbf{electronically} in real time. As a result, when a beam of electromagnetic (EM) wave impinges on the surface, the \textit{phase, amplitude, or frequency} of the reflected wave can be controlled and \textbf{programmed} in a software-defined way using precise mathematics~\cite{chen2019intelligent,tang2019wireless}. Figure~\ref{fig:irs_arc} shows how IRS can help reduce wireless transmission energy cost of IoTs. In the first use case, the IRS can simply enhance the functionalities and capabilities of ambient backscattering communication in EH-IoTs. In the second use case, IRS can be deployed in the environment, such as coated over a wall, to control the wireless channel and achieve superior signal-to-noise ratio at the receiver as a result. As such, the IoT transmitter, which may still use the conventional generative radio, can afford to use reduced transmission without sacrificing signal-to-noise-ratio at the receiver. Table \ref{t:evolution} summarizes the three phases of backscatter evolution, while we survey the recent achievements of IRS-enabled wireless communications in the following section.

\begin{table*}[t]
\centering
\caption{Recent advancements in IRS-assisted wireless communications.}
\label{tab:metasurface summary}
\setlength\tabcolsep{2.5pt}
\ra{1.2}
\begin{tabular}{@{}cll@{}}\toprule
   
\textbf{ Context} &\textbf{Problem/Objective}  &\textbf{Methods}   \\ \midrule

\multirow{2}{*}{Modeling} &capacity upperbound study~\cite{hu2018beyond}    & Theoretical analysis of received signal model at IRS and achievable capacity. \\ \cmidrule{2-3}

& IRS potential study~\cite{liaskos2019novel,liaskos2018realizing}    & Simulation based on full 3D ray-tracing to demonstrate wireless and security potential. \\ \cmidrule{1-3}

\multirow{12}{*}{Low-energy} &maximize energy efficiency~\cite{huang2018energy,huang2019reconfigurable}   & Joint optimization using gradient descent and sequential fractional programming. \\ \cmidrule{2-3}
 
\multirow{12}{*}{ MIMO}&maximize total received SNR~\cite{wu2018intelligent}   &Proposed a distributed algorithm using alternating optimization. \\ \cmidrule{2-3}

\multirow{12}{*}{using IRS}&maximize spectral efficiency~\cite{yu2019miso} &  Proposed two algorithms based on fixed point iteration and manifold optimization.   \\ \cmidrule{2-3}

&maximize secrecy rate~\cite{yu2019enabling}   &Proposed two algorithms based on block coordinate descent and minorization maximization. \\ \cmidrule{2-3}

&maximize secrecy rate~\cite{xu2019resource}   &Proposed an algorithm based on alternating optimization and semidefinite relaxation.  \\ \cmidrule{2-3}

&interference canceling~\cite{tan2016increasing} &Optimally adjusting the phase shifts of IRS to achieve high spectrum-spatial efficiency. \\ \cmidrule{2-3}

&minimize Tx power at AP~\cite{wu2019beamforming,wu2019beamforming2}   &Proposed two algorithms based on successive refinement and ZF based linear precoding. \\ \cmidrule{2-3}

&maximize weighted sum-rate~\cite{guo2019weighted}   & Joint optimization using fractional programming and other three proposed algorithms. \\ \cmidrule{2-3}

&cascaded channel estimation~\cite{he2019cascaded}    & Proposed a two-stage algorithm including matrix factorization and matrix completion. \\ \cmidrule{2-3}

&cascaded channel estimation~\cite{taha2019enabling}    &Proposed an approach based on compressive sensing and deep learning. \\ 
\cmidrule{1-3}

 &transceiver design~\cite{tang2019wireless}   & Implemented QPSK IRS backscattering achieving 2.048 Mbps video streaming. \\ \cmidrule{2-3}
 
IRS Backscattering&transmitter design~\cite{tang2019programmable}   & Implemented 8PSK  IRS backscattering achieving 6.144 Mbps at 4.25 GHz.\\ \cmidrule{2-3}

&efficient control of meta-surface~\cite{petrou2018asynchronous}   & Proposed to embed synchronous digital circuit to increase adaptability of meta-atoms.  \\

 \bottomrule 

\end{tabular}
\end{table*}

\subsubsection{Recent Advancements in IRS-based Wireless Communications}
The recent advancements in IRS-based wireless communications can be classified into the following three categories (they are summarized in Table~\ref{tab:metasurface summary}):

{\bf Modelling and analysis of IRS:} Using tools such as ray-tracing, researchers have developed analytical models to study the theoretical capacity bounds of IRS under different deployment settings and constraints~\cite{hu2018beyond,liaskos2019novel,liaskos2018realizing}. The derived models can be used to gain insight to the capabilities of IRS for many envisaged IoT communications scenarios under specified power budget.

{\bf Low complexity and low energy MIMO with IRS:} Although multiple-input multiple-output (MIMO) technology
has significantly improved the spectrum and energy efficiency
of wireless communication systems, it suffers from high hardware complexity and power consumption. Recently, researchers have shown~\cite{huang2018energy,huang2019reconfigurable,wu2018intelligent,yu2019miso,yu2019enabling,xu2019resource,tan2016increasing,wu2019beamforming,wu2019beamforming2,guo2019weighted,he2019cascaded,taha2019enabling} that using IRS, it is possible to significantly improve the performance of MIMO, which ultimately translates to reduced number of transmitting and receiving antennas as well as reduced power consumption for both transmitters and receivers. These solutions will help EH-IoTs to achieve high MIMO gain under tight power constraints.

{\bf Ambient backscattering with IRS:} Researchers have successfully designed hardware circuits and control algorithms~\cite{tang2019wireless,tang2019programmable,petrou2018asynchronous} to realize ambient backscattering with IRS. These systems allow an IRS surface to realize advanced modulations, e.g., Quadrature Phase Shift Keying (QPSK)  and 8PSK, over the reflected signals to realize video transmissions at high data rates.

\subsection{Lessons Learned}

The key lessons learned from the recent studies in energy harvesting communications are as follows:

\begin{itemize}
\item Reinforcement learning-based transmission optimization is proving very effective for practical scenarios where prior knowledge of channel states and energy arrivals is limited.

\item Packet-less communication has been shown to be effective in detecting and monitoring certain types of events with minimal power consumption. However, the current applications of packet-less communication are limited and most works are based only on theory. There exist future research opportunities in this space to realize practical packet-less communication systems.

\item The reflective radio technology, especially those based on the emerging meta-surfaces, exhibits a huge potential in dramatically cutting the transmission power cost of future EH-IoTs. However, the IRS technology is in its infancy at the moment and there exist significant future research opportunities to pursue in this space.

\item For IRS-enabled wireless communications, the researchers have predominantly used simulations to solve various optimization problems, while actual prototyping and experiments were rare.

\end{itemize}

\section{Future Research Directions}
\label{section:futurework}

\subsection{Improving Context Sensing from EH Patterns}
The lesson that we have learned from Section \ref{section:sensing} is that energy harvesters could potentially replace specialized sensors in EH-IoTs, thus saving significant power consumption, but at the cost of reduced context detection performance. To improve the sensing performance, novel research is required on both algorithm and hardware design.

\subsubsection*{Algorithm}
From algorithms perspective, the application of deep learning has shown great promise for many challenging detection problems, such as speech detection, face recognition, natural language processing, and so on. However, its application to context detection from energy harvesting signals have not been explored yet. Once EH-IoTs are deployed widely, they will generate a huge amount of data, which will create the opportunity to apply deep learning to improve context detection accuracy. In the meantime, simulations could be used to generate the required data for training. 

\subsubsection*{Hardware}
From hardware perspective, more advanced and refined energy harvesters are required. Most EH-based context sensing research to date used the very basic kinetic energy harvester (KEH) or solar cell hardware, which have certain limitations for context recognition. For example, the frequency response band of most KEHs is narrow and well-tuned to maximize their energy harvesting efficiencies for target applications. As a result, such KEH hardware is more sensitive to capture the signal within its resonance band, while responding less to the signals outside their narrow response band. On the contrary, specialized sensors like accelerometers and microphones are engineered with a wide and flat response curves ranging from 1Hz to several thousands of Hz over the frequency of interests. Also, accelerometers can provide measurements in three dimensions, while the popular KEH devices only harvest energy from a single direction.

A new research direction would be to explore more advanced KEH hardware, such as those that can harvest energy from multiple axes~\cite{aktakka2015}, multiple elements within an array~\cite{array}, or multiple modes such as from both piezoelectric and electromagnetic effects~\cite{chen2015piezoelectric}. Similarly, the use of more advanced solar cells which can harvest energy from a wider range of optical frequencies~\cite{umetsu2019ehaas} may help detect contexts more accurately.

Currently, energy harvesting hardware is designed purely with the objective of improving the energy harvesting density or efficiency of the product without any consideration of its context detection capability. A potential multidisciplinary research direction would be to explore new materials and processes that can jointly optimize both energy harvesting capacity and context detection performance.  

Finally, existing literature so far explored context sensing using a single mode of energy harvesting. In future, IoTs may combine multiple modes of energy harvesting, such as combine both kinetic and solar, to boost the power supply. Such multi-modal EH-IoTs would provide richer EH signals, which could be exploited for more accurate context detection. Designing algorithms that can effectively fuse information from multiple EH signals would also be an interesting future research direction.

\subsection{Context Sensing from RF EH}
Section III revealed that, compared to kinetic and solar EH based context sensing, research on RF EH based sensing is rare. Recently, the received signal strength (RSS) and channel state information (CSI) of Wi-Fi signals have been extensively exploited to perform a multitude of sensing applications, such as activity recognition~\cite{wang2017device}, heartbeat monitoring~\cite{wang2017tensorbeat}, fall detection~\cite{wang2017wifall}, occupancy counting~\cite{depatla2015occupancy}, gait recognition~\cite{wang2016gait} and many more. Since Wi-Fi signal is also a potential source of energy in RF EH, sensing from the EH patterns from Wi-Fi signal should be possible. However, this branch of research is yet unexplored in the literature and therefore new research is needed to integrate RF EH with context sensing to reduce power consumption of RF sensing.

\subsection{Privacy Leakage of Energy Harvesting}
As energy harvesting patterns have been demonstrated to be able to infer various private user contexts like gait and activity (see Section \ref{section:sensing}), a malware with access to power generation data could cause serious privacy concerns for the user. Even legitimate apps can pose such privacy risks by accessing power data, which is possible because users often do not pay attention to all types of sensor data access requests during the app installation~\cite{liang2018deep}.

Techniques such as perturbation, encryption, anonymization, differential privacy, and information-aware privacy have been extensively exploited to protect user privacy in various scenarios~\cite{xiao2018information,raval2019olympus}. Most recently, deep learning based privacy-preserving has been investigated as well~\cite{malekzadeh2019mobile}. In the future, novel privacy preserving schemes should be explored for energy harvesting data. For example, access control or authentication towards power data and deep leaning based energy harvesting data perturbation can be potential future research directions. 

\subsection{Hardware Assistance for Intermittent Computing}
During software development, program debugging is highly recommended to ensure an IoT device can operate correctly in real deployments. For example, a Devpack is needed when debugging a TI SensorTag \cite{sensortag}. However, conventional debugging cannot fully address the concerns of EH-IoTs. First, it provides constant power to the target device, which would treat an EH device as a grid- or battery-powered device. Therefore, bugs that may occur under energy harvesting conditions may not be detected. Second, developers usually insert some additional codes (e.g., \textit{printf()}) to monitor the state of the program during debugging. But the execution of these codes consumes extra energy and may incorrectly lead to the Sisyphean problem as discussed in Section \ref{section:computing}. Work on designing new debugging concepts and tools for EH-IoTs is limited with the exception of~\cite{colin2016energy}, which attempts to actively manipulate the amount of energy delivered to the target device in order to account for any additional energy consumption due to extra debugging codes. More research on energy-aware debugging is required to realize highly efficient and reliable debugging and development support for EH-IoTs.

\subsection{Secure Communication for EH-IoTs}
Due to low and unpredictable energy supply, EH-IoTs can be more vulnerable to security attacks compared to the conventional battery-powered IoTs. For example, as an EH-IoT may suspend in the middle of secure communication protocols, it opens new attack horizons, e.g., DoS attack to the gateway that communicates with the EH-IoT device. Due to low power requirements, backscattering is considered an attractive communication option for EH-IoTs. However, the passive operation makes backscatter communications vulnerable to various security threats like eavesdropping and jamming~\cite{van2018ambient}. While most of the security attacks can be effectively addressed by encryption, the overhead of sharing secret keys make them challenging for EH-IoTs that must operate with minimal energy supply. A future research direction could be to exploit the energy harvesting signals for generating keys dynamically with minimal power consumption. For example, in~\cite{lin2019h2b}, it has recently demonstrated that the tiny vibrations generated by human heartbeats can be measured by piezoelectric-based wearable KEH, which in turn can be used to generate symmetric keys for two IoTs worn by the same user (both devices are subject to the same heartbeats). However, whether EH signals like photocurrents from solar cells, can also be useful for key generation for non-wearable IoTs remains to be explored.

\subsection{Wireless Protocols for EH Communications}
To cope with the limited amount of harvested energy, recent developments in EH communications usually require the EH devices to operate in duty-cycling mode and dynamically adjust the duty cycles based on real-time energy arrivals. Unlike battery-powered duty-cycling that can wake up a node at any time, EH-IoTs cannot wake up unless a certain amount of energy is accumulated, which poses challenges on the design of MAC protocols. For example, the synchronization between the transmitter and receiver becomes difficult as it requires both devices to have enough energy at the same time, and the utilization efficiency of the channel will be low. 

In addition, such duty-cycling scheme introduces difficulties in the design of routing protocol in multi-hop EH networks. Due to the desynchrony of energy availability in the EH nodes, the relay selection and routing overhead in terms of energy utility~\cite{jia2017joint}, end-to-end delay, and overall throughput should be incorporated in the routing protocol design.

\subsection{Intelligent Reflective Surface Communication for IoT}
Although the IRS-based ambient backscatter communication has been successfully realized to achieve data rates in the order of several Mbps~\cite{tang2019wireless,tang2019programmable}, these works are just initial attempts to verify the concept without exploiting the full programmability of IRS. Specifically, the reflection property of each IRS element can be controlled independently in principle, while the current experiments merely controlled all elements on the surface with the same voltage, thereby losing many degrees of freedom for modulation. Thus, more advanced circuits and algorithms are required for optimal control of all the elements individually. Such optimizations are likely to be very complex to solve, and hence reinforcement learning or deep learning based IRS control designs can be promising future directions. The concept of IRS allows manipulation of the amplitude, frequency, and phase of the reflected signal, while existing research modulates only the phase. As such, a future direction would be to explore more advanced and high-order modulation schemes like QAM (Quadrature Amplitude Modulation) and OFDM. Last but not the least, more large scale deployments of IRS and data collection/analysis from such deployments would be an important future direction to gain real insights to the practical benefits of this emerging technology.

\subsection{Integrated System Design for EH-IoTs}
As shown in Figure~\ref{fig:architecture}, current EH-IoT architecture simply uses an EH module to replace the battery, which can enable wide and rapid adoption of energy harvesting without requiring significant hardware modification in the IoT design. However, given the preciousness of harvested energy, this loose architecture fails to optimize the energy utilization efficiency. Thus, a future research direction can be a more holistic design, where the energy harvester, energy storage, microcontroller, as well as the radio module can be integrated together. Moreover, to connect these components tightly, an advanced power management circuit should be customized to avoid energy loss due to hardware imperfections. For example, Texas Instruments has released some power management modules and guidance specifically for energy harvesting~\cite{tipowerma}.

In addition, in RF energy harvesting communications, the transceiver design that optimizes both data reception and energy harvesting remains an open research problem. Typically, the energy conversion efficiency from RF signals to direct current (DC) depends on the density of harvested energy and such efficiency is low when the amount of harvested energy is small. Thus, with the premise of satisfying communication, more efforts are required to improve the energy conversion efficiency from hardware perspective, e.g., more sophisticated circuit design.

\subsection{Practical Implementations and Deployments of EH-IoTs}
While powering IoTs using energy harvesting has been proposed for a while, most of the research efforts is based on simulation or a single prototype. Several important challenges arise in large-scale practical implementations and deployments of EH-IoTs. First, most EH-IoTs are supposed to be manufactured with low-cost, small form factor, self-autonomy, robust operation, and long-term lifetime. As a result, the target applications can be achieved with high-efficiency while involving minimum human intervention. Thus, compact EH-IoT design and fabrication is required. Second, the amount of harvested energy might not be adequate or stable enough to support the operation of EH-IoTs. Thus, more advanced techniques for energy harvesting should be explored. For example, in RF energy harvesting, the popularity of using multiple antennas for enhanced communication performance also creates the opportunity to improve EH performance. Multiple antenna technologies like
beamforming, distributed antennas, and massive MIMO can be employed to increase the amount of harvested energy. In addition, heterogeneous energy harvesting with multiple energy sources is another way to compensate the energy scarcity. Third, a simulation or testing stage is usually required before any large-scale field deployment to avoid the overhead of re-deployment in case of bugs. However, due to the uniqueness and dynamics of energy arrival depending on different energy sources, it is difficult to emulate actual EH conditions. Therefore, development tools like Ekho~\cite{hester2014ekho}, which can record physical EH condition and recreate such condition in the lab environment, and systematic testing methodology require further investigation in the future.  

\section{Conclusion}
\label{section:conclusion}

Energy harvesting is a promising new approach to perpetually power a growing number of IoT sensors. The key challenge for EH-IoTs is to ensure smooth operation over unpredictable power supply, which requires a holistic design of sensing, computing, as well as communications. In recent years, various approaches and solutions have been proposed to optimize the use of harvested energy. In this paper we have classified, compared, and analyzed such solutions, identified the lessons learned, and discussed potential future directions for research. We have also identified and summarized relevant standards activities that are currently being undertaken to promote interoperability and accelerate the deployment of future EH-IoTs.

\balance

\bibliographystyle{IEEEtran} 
\bibliography{ref}

\end{document}